\documentclass[10pt,journal,compsoc]{IEEEtran}

\ifCLASSOPTIONcompsoc
  \usepackage[nocompress]{cite}
\else
  \usepackage{cite}
\fi
\ifCLASSINFOpdf
\usepackage{cite}
\usepackage{amsmath,amssymb,amsfonts}
\usepackage{algorithmic}
\usepackage{graphicx}
\usepackage{textcomp}
\usepackage{graphicx}
\usepackage{subcaption}
\usepackage{multirow}
\usepackage{mwe}
\usepackage{booktabs}
\usepackage{mathrsfs}
\usepackage{epstopdf}
\usepackage{epstopdf}
\DeclareGraphicsExtensions{.png,.pdf}

\usepackage{booktabs}

\DeclareMathOperator*{\argmin}{argmin}

\newcommand{\bigzero}{\mbox{\normalfont\Large\bfseries 0}}

\begin{document}
%
\title{A Multi-View Discriminant Learning Approach for Indoor Localization Using Bimodal Features of CSI}
%
%
%
%

\author{Tahsina Farah Sanam,~\IEEEmembership{Student Member,~IEEE}, Hana Godrich,~\IEEEmembership{Senior Member,~IEEE}

\IEEEcompsocitemizethanks{\IEEEcompsocthanksitem The authors are with the Department of Electrical and Computer Engineering, Rutgers, The State University of New Jersey, New Brunswick, NJ, 08854, USA. ( e-mail: tahsina.farah@rutgers.edu; godrich@soe.rutgers.edu)}
\thanks{Manuscript received August 10, 2019; revised xxx xx, 2019.}}

\IEEEtitleabstractindextext{%
\begin{abstract}
With  the  growth  of  location-based  services,  indoor  localization is attracting great interests as it facilitates further ubiquitous environments. Specifically, device free localization using wireless signals is getting increased attention as human location is estimated using its impact on the surrounding
wireless signals without any active device tagged with subject. In this paper, we propose MuDLoc, the first multi-view discriminant learning approach for device free indoor localization using both amplitude and phase features of Channel State Information (CSI) from multiple APs. Multi-view learning is an emerging technique in machine learning which improve performance by utilizing diversity from different view data. In MuDLoc, the localization is modeled as a pattern matching problem, where the target location is predicted based on similarity measure of CSI features of an unknown location with those of the training locations. MuDLoc implements Generalized Inter-view and Intra-view Discriminant Correlation Analysis (GI$^{2}$DCA), a discriminative feature extraction approach using multi-view CSIs. It incorporates inter-view and intra-view class associations while maximizing pairwise correlations across multi-view data sets. A similarity measure is performed to find the best match to localize a subject. Experimental results from two cluttered environments show that MuDLoc can estimate location with high accuracy which outperforms other benchmark approaches.
\end{abstract}

}

\maketitle

\IEEEdisplaynontitleabstractindextext

%
\IEEEpeerreviewmaketitle

\IEEEraisesectionheading{\section{Introduction}\label{sec:introduction}}

%
%
%
%
\IEEEPARstart{L}{earning} important human contextual information is one of the fundamental features to establishing a smart environment. The ability to localize various subjects indoor can potentially support a broad array of applications including elder care, rescue operations, vehicle parking management, building occupancy statistics, security enforcement, etc. Unlike outdoor localization that can rely on the use of Global Positioning System (GPS), that is based on transmission of Line-of-Sight (LOS) paths, indoor localization suffers from a lot of challenges due to indoor radio propagation, such as multipath, fading, shadowing, etc. \cite{why_IL1}. 
Wireless signals, specifically Wi-Fi signals have emerged as one of the most pervasive signals for this application. Human presence is interfering with these signals. By observing the channel features over time, people's location can be inferred by comparing them against pre-constructed signal profiles, which is commonly known as fingerprinting approach \cite{RSS_to_CSI}. Most fingerprinting-based localization approaches rely on coarse-grained Received Signal Strength (RSS) \cite{Radar,horus, Nuzzer,Xu_mob_com}. These vary over distance on the order of the signal wavelength and fluctuates over time, resulting in localization with lower accuracy \cite{RSS_to_CSI}.To improve localization accuracy, fine grained PHY layer CSI has recently gained significant attention for different applications \cite{CSI_Virtual_Reality_Devices,Healthcare,WiFall}, which is available in several Wi-Fi network interface cards (NIC) \cite{Halperin, Halperin1}. Unlike RSS, in IEEE 802.11n communication, Multiple Input Multiple Output (MIMO) OFDM systems provide CSIs with amplitude and phase for subcarrier level channels for each antenna link. CSI is richer in multipath information and more stable than RSS for a given location, hence, a preferable choice to realize an improved indoor localization system \cite{deep,FIFS,FILA,CSI-MIMO,Sanam1,DisLoc,Pilot,LiFS}.

For most existing indoor localization methods, the subject being tracked is assumed to carry mobile devices that is used for localization \cite{deep,FIFS,FILA,CSI-MIMO}. However, accidental or intentional detachment of the device from the subject can terminate the tracking process. In wireless signal-based device
free localization, the subject being tracked does not need
to carry any device, instead the feature pattern of
the wireless signal that is being interfered by the
presence of the subject is utilized for estimating the location. To
this end, indoor localization through the pattern matching
approach has emerged as an effective technique to facilitate for device free indoor localization systems, where the location of a subject is predicted
based on the similarity measure of the CSI features of the
unknown location with those of the training locations. 

Various device free indoor localization methods have been explored in \cite{Pilot,LiFS,Defi,why_wifi1,FreeSense,DeviceFree_RSS_1,DeviceFree_RSS_2, Xu_mob_com}. These methods consider either RSS value of wireless signals or only the amplitude of CSI. As such, lots of useful information embedded with the phase is not used. In \cite{deep, FIFS}, authors perform CSI based location estimation using Wi-Fi enabled device tagged with the subject. In \cite{Xu_mob_com}, the device free localization is performed using probabilistic classification approaches that are based on discriminant analysis of wireless sensor-based RSS value. Device free indoor localization using CSI has been explored in \cite{Pilot,LiFS}. In \cite{Pilot}, authors perform CSI based device free localization through a probabilistic approach and showed improvement in localization with $85\%$ accuracy where \cite{LiFS} adopts a power fading model-based localization and achieves $90\%$ accuracy with 1.5 meter localization error. Both methods considered only the amplitude values of CSI from a single AP.  
Moreover, all these methods consider CSI measurements collected from multiple Access Points (AP) independently. However, a set of CSI measurements,
simultaneously recorded from multiple APs for a particular target location should share some common features, which might be
correlated. Some useful information, involved with multiple
OFDM channel correlations, may lost if measurements from
each AP are considered independently. Therefore, joint
utilization of CSI measurements from multiple APs can be used
to achieve higher accuracy for location estimation.

In this paper, MuDLoc, a multi-view discriminant learning approach for device free indoor localization
using CSI is proposed. This method utilizes CSI
measurements recorded from multiple APs for a particular target
location in order to extract common features shared by all
APs. However, direct matching of the data samples across
various feature spaces is infeasible. Subspace learning offers an effective approach to solving the problem, which learn a common feature space from multi-view spaces. Therefore, MuDLoc develops a multi-view learning approach to extract joint spatial features from CSI measurements recorded from multiple APs. Various multi-data processing techniques have been reported in literature. Canonical Correlation Analysis (CCA) is one of the multi-data processing methods that deals with linear relationship between two or more multidimensional variables \cite{hafiz_ref2,hafiz_3, DCE_haghihat,feature_fusion1,feature_fusion2}. Multi-view CCA (MCCA) was developed as an extension of CCA to find multiple linear transforms that maximize overall correlation among canonical variates from multiple sets of random variables \cite{MCCA,MCCA_ref1,KETTENRING}. However, MCCA does not take discriminant information into account, which may degrade classification performance across classes. The supervised information was incorporated in a generalized multiview analysis framework (GMA), leading to a discriminant common subspace \cite{GMA}. However only the intra-view discriminant information was considered in GMA, ignoring inter-view discriminant information, which may degenerate performance of cross-view matching. Adopting multi-view strategy for cross-view recognition, the proposed MuDLoc method implements Generalized Inter-view and Intra-view Discriminant Correlation Analysis (GI$^{2}$DCA), a subspace learning approach that can learn single unified discriminant common space from the joint spatial filtering of multiple sets of CSI data recorded for a particular target location. In this common space, the between-class variations from both inter-view and intra-view are maximized, while keeping the projections of different views close to each other in the latent common space. Therefore, both inter-view and intra-view class structures are preserved. Moreover, the system exploits not only the amplitude of CSI, but also the phase information of multi-view CSI data to learn the discriminative common space. The phase difference for two receiver antennas is more stable with 5GHz Intel 5300 NIC \cite{phaser_biloc19,pi_splicer}. In addition to amplitude information, the proposed method utilizes CSI phase difference information between adjacent antennas for consecutive packets under the multi-view setting. This enables the exploitation of  complete multipath features to achieve a higher localization accuracy. Once the discriminant features from multi-view CSI are obtained, the system shifts the localization problem to a cell/grid classification problem by employing a simple but efficient euclidean distance-based similarity measure approach. It finds the best cell match to localize a test subject. The main contributions of the paper can be summarized as follows. 
\begin{figure}[t]
\centering
\includegraphics[width=3.3in]{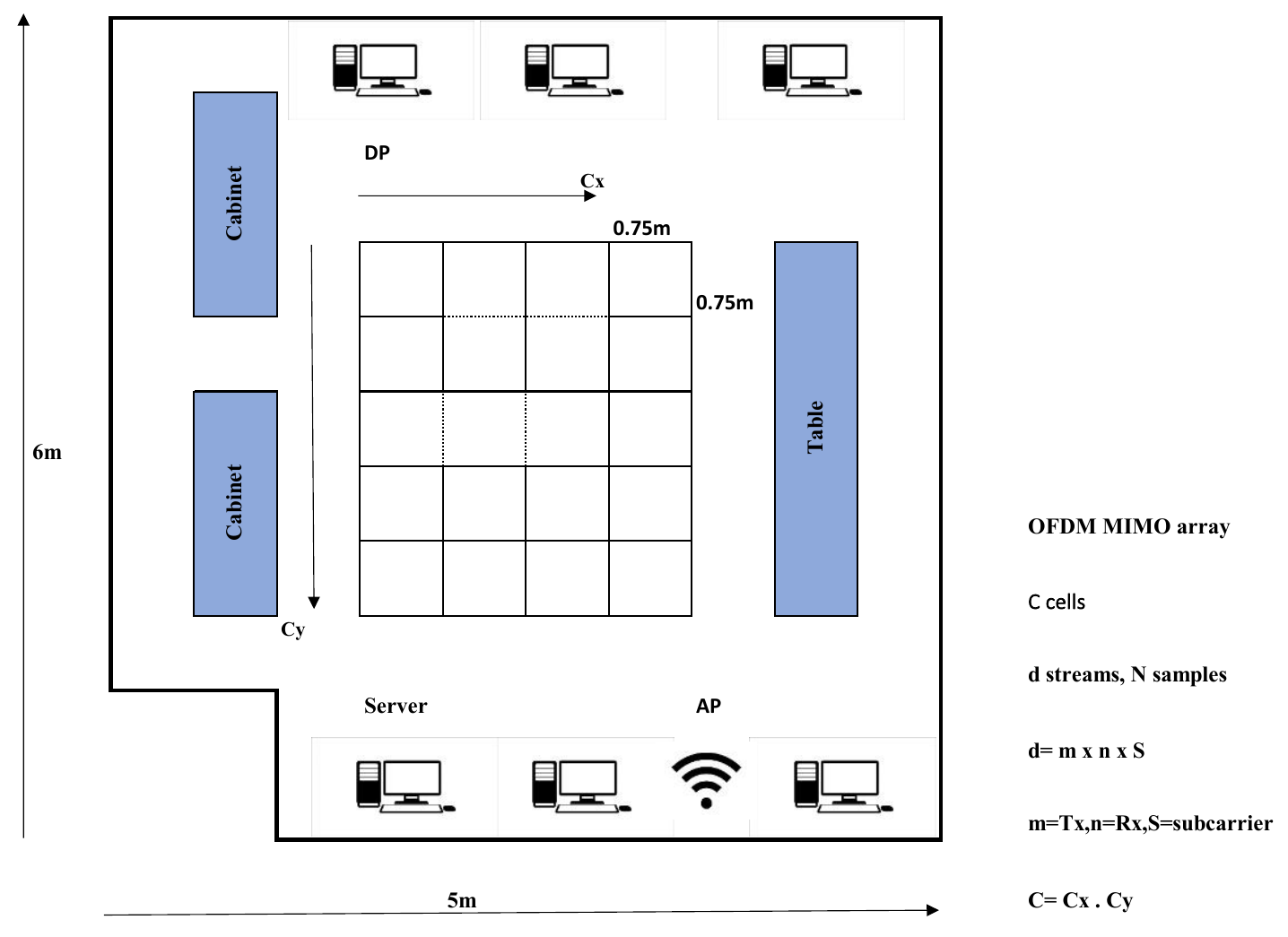}
\caption{System Model}
\label{system}
\end{figure}
\begin{enumerate}
    \item Utilizing CSI measurements from multiple AP through the multi-view subspace learning approach using GI$^{2}$DCA, where both inter-view and intra-view class structures are preserved. 
    
    \item The MuDLoc system implements multi view learning of CSI by leveraging both the amplitude and phase difference of adjacent antennas in a MIMO OFDM system, and thereby complete multipath features are utilized in order to achieve higher localization accuracy.
    
    \item Extensive experimentation performed in two
    cluttered indoor environments are used to verify the
    effectiveness of MuDLoc, demonstrating it outperforms
    previously proposed state-of-the-art localization methods. 
\end{enumerate}

The rest of the paper is structured as follows. Section \ref{sec: motivation} presents the motivation behind the proposed MuDLoc system. Preliminaries on CSI, phase information and basic multi-data processing using CCA are described in Section \ref{sec:preli}. The MuDLoc method, a multi-view discriminant learning approach for indoor localization, is introduced in Section \ref{sec: proposed}. Section \ref{sec:Experimental Study} describes the
system experimental setup and evaluates the
performance of the proposed method. Finally, concluding
remarks are discussed in Section \ref{sec:Conclu}.

 
\section{Motivation}\label{sec: motivation}
The proposed MuDLoc system consists of three basic hardware elements in a WLAN infrastructure: access points (AP), detecting points (DP) and a server. Each pair of AP and DP establishes a radio frequency (RF) link. Beacon messages are broadcast periodically by the APs. A WiFi compatible device is used as the DP that interacts with the APs and the server. Once the beacon message is received, the DP records the raw PHY layer CSIs across multiple subcarriers from the multiple APs (views) and sends them to the server to store and process. The area is considered as a grid of small square cells and there are \textit{C} cells in that area of interest as shown in Fig. \ref{system}. Our goal is to use CSI fingerprints collected at Detecting Points (DP) from multiple APs in order to classify a testing entity with an unknown cell ID. 
\begin{figure}[t]
\centering
\includegraphics[width=3in]{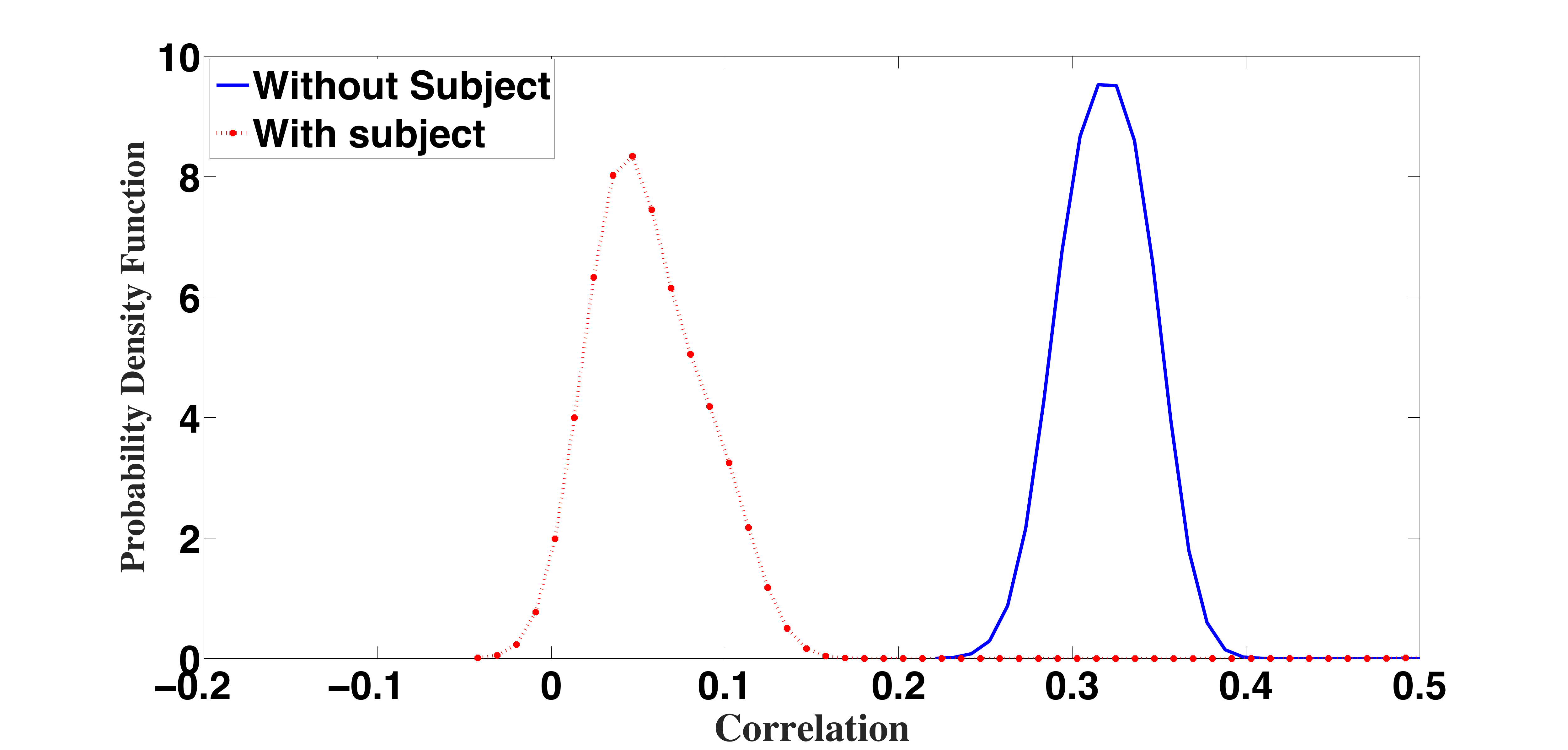}
\caption{Effect of Subject Appearance on CSI feature Shift.}
\vspace{-0.2cm}
\label{anomaly}
\end{figure}
\begin{figure}[!t]
\centering
\includegraphics[width=3in]{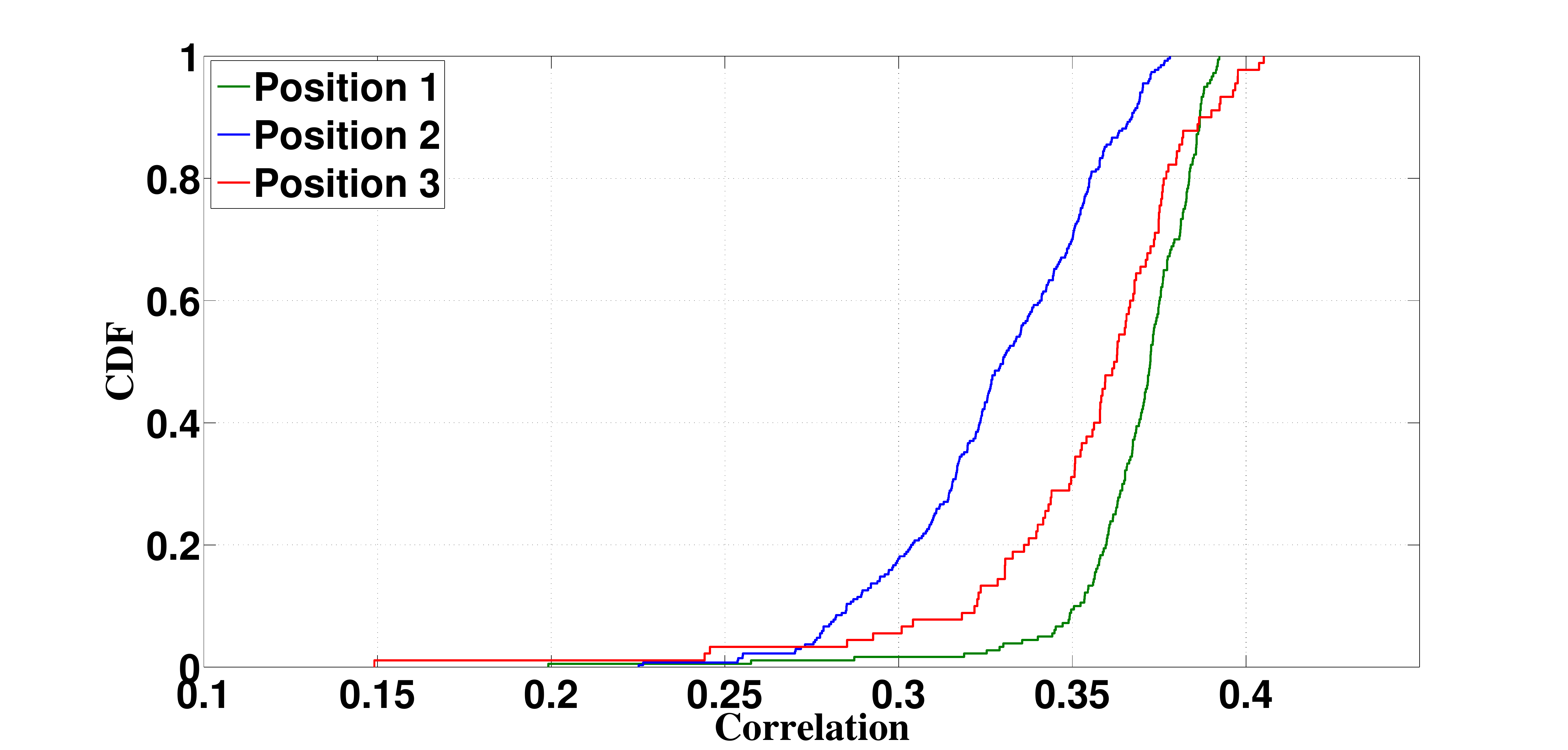}
\caption{Location-Specific CSI feature Variance.}
\vspace{-0.4cm}
\label{location_feature}
\end{figure}
Our initial motivation to utilize CSI for device free localization stems from several observations. CSI over MIMO channel can reveal the change in the environment due to the appearance of an entity. Fig. \ref{anomaly} shows that there is a shift in the feature pattern of empirical probability distribution function of CSI correlation when a subject appears in an empty room. Moreover, CSI from MIMO channel can also identify entity at different location, as it reveals different feature pattern at different location as depicted in Fig. \ref{location_feature}. This is because in MIMO-OFDM systems, for each transmit and receive antenna pair, CSIs over multiple subcarriers suffer from different scattering due to multi-path. This work exploits these two findings on CSI in MIMO-OFDM system to generate CSI fingerprint for each location. 

The motivation for utilizing multi-view learning of CSI measurements stems from the idea that, people can see a location differently. Similarly, CSI data for a particular location can be represented with various point of views (different locations AP measurements) and with different modalities (amplitude and phase information). Specifically, amplitude and phase information of CSI measurements, extracted from multiple APs located at different locations reflect different characteristics of the multipath patterns, affected by the presence of a subject at a particular location. Consequently, some common features contained in the several real time CSI measurements from multiple APs could be more useful for location estimation of a test subject in contrast to that of a single AP (view) based localization approach. Motivated by this idea, MuDLoc exploits CSI measurements recorded from multiple location APs through joint spatial filtering in order to utilize a multi-view learning approach for better localization performance.
\begin{figure}[t]
        \centering
        \begin{subfigure}[b]{0.2\textwidth}
            \centering
            \includegraphics[width=\textwidth]{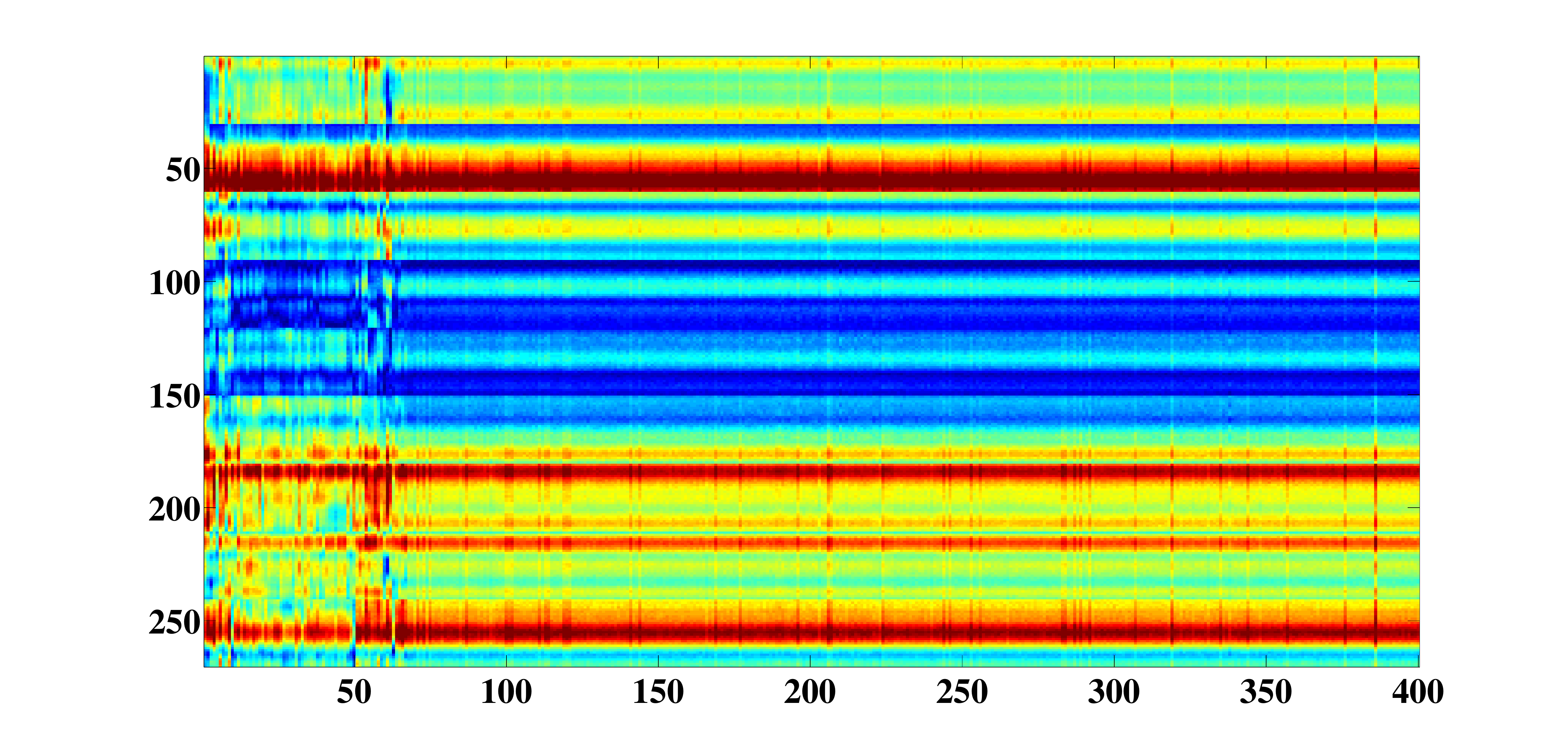}
            \caption[Network2]%
            {{\small Location 1, AP1}}    
        \end{subfigure}
        \hfill
        \begin{subfigure}[b]{0.2\textwidth}  
            \centering 
            \includegraphics[width=\textwidth]{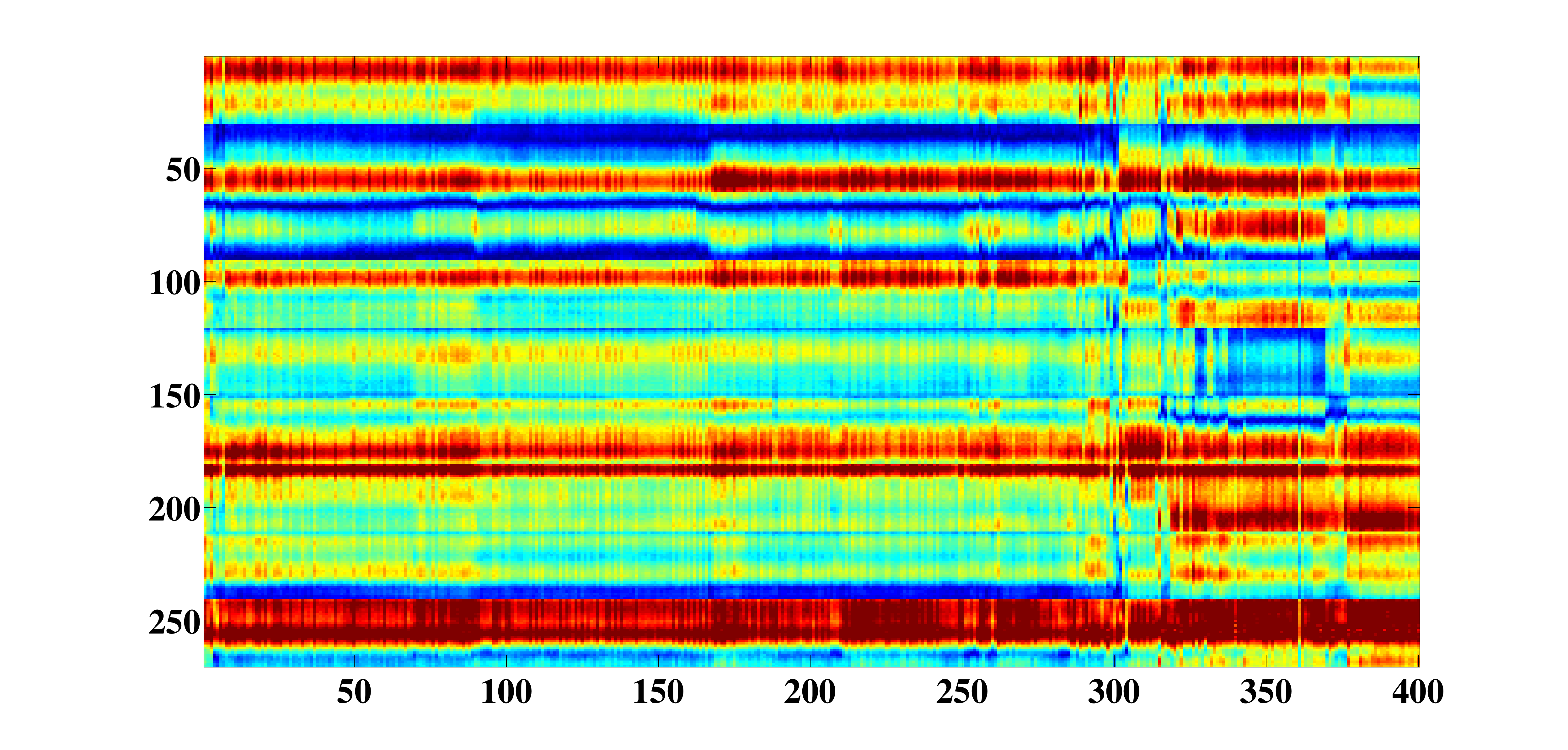}
            \caption[]%
            {{\small Location 1, AP2}}    
        \end{subfigure}
        \vskip\baselineskip
        \begin{subfigure}[b]{0.2\textwidth}   
            \centering 
            \includegraphics[width=\textwidth]{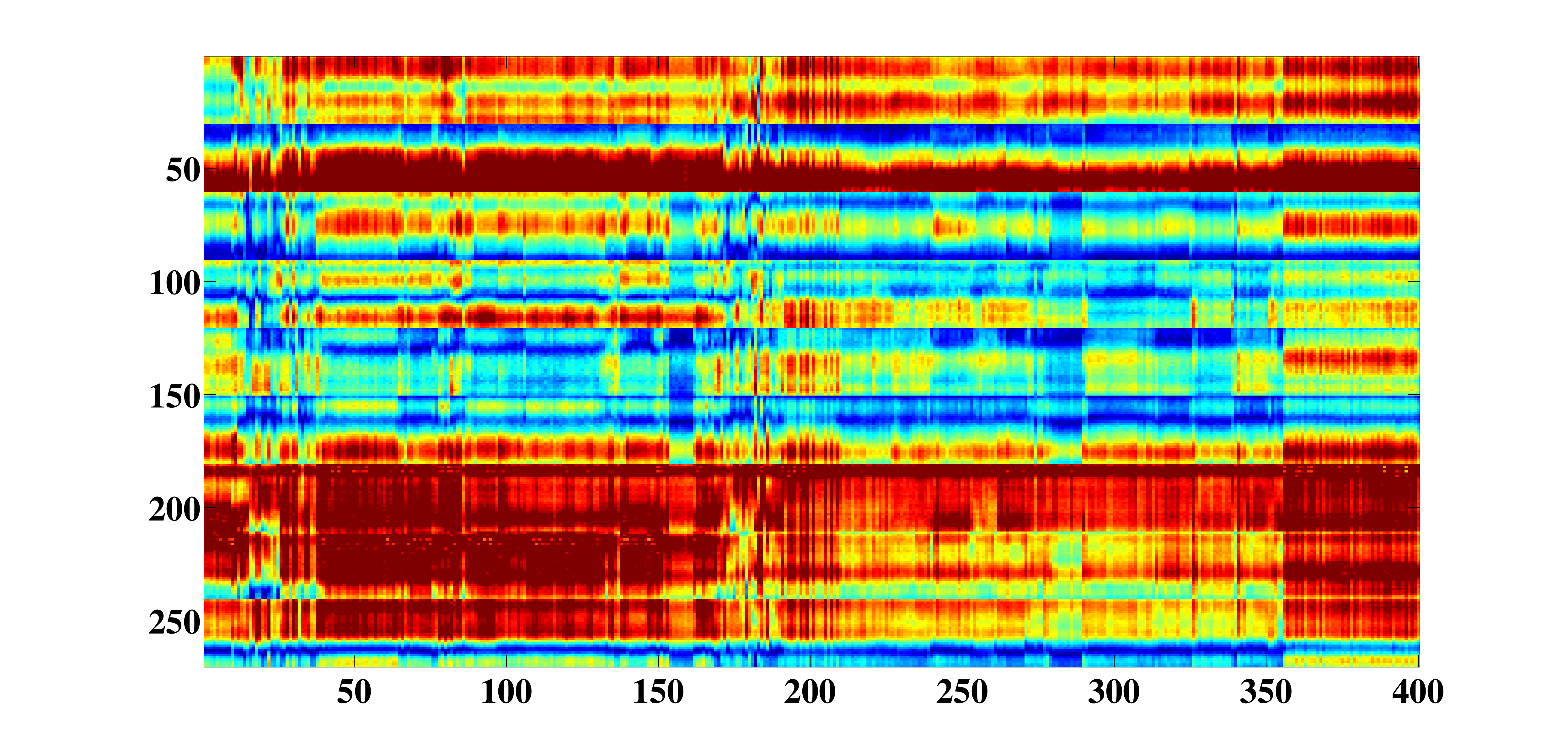}
            \caption[]%
            {{\small Location 2, AP1}}    
        \end{subfigure}
        \hfill
        \begin{subfigure}[b]{0.2\textwidth}   
            \centering 
            \includegraphics[width=\textwidth]{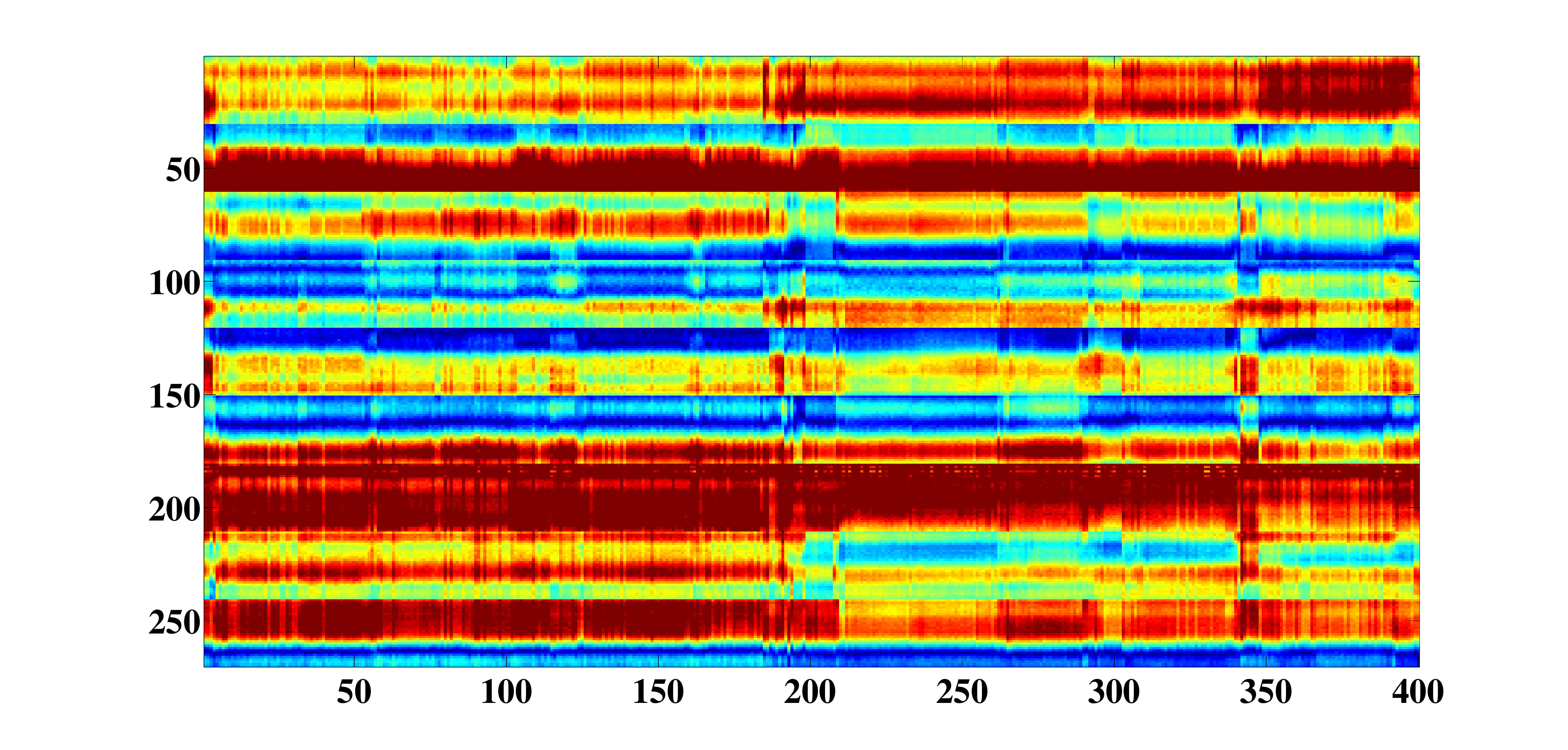}
            \caption[]%
            {{\small Location 2, AP2}}    
        \end{subfigure}
        \caption{Feature images of different locations using CSI amplitude.}
        \label{CSI_image_amp}
    \end{figure}

\begin{figure}[t]
        \centering
        \begin{subfigure}[b]{0.2\textwidth}
            \centering
            \includegraphics[width=\textwidth]{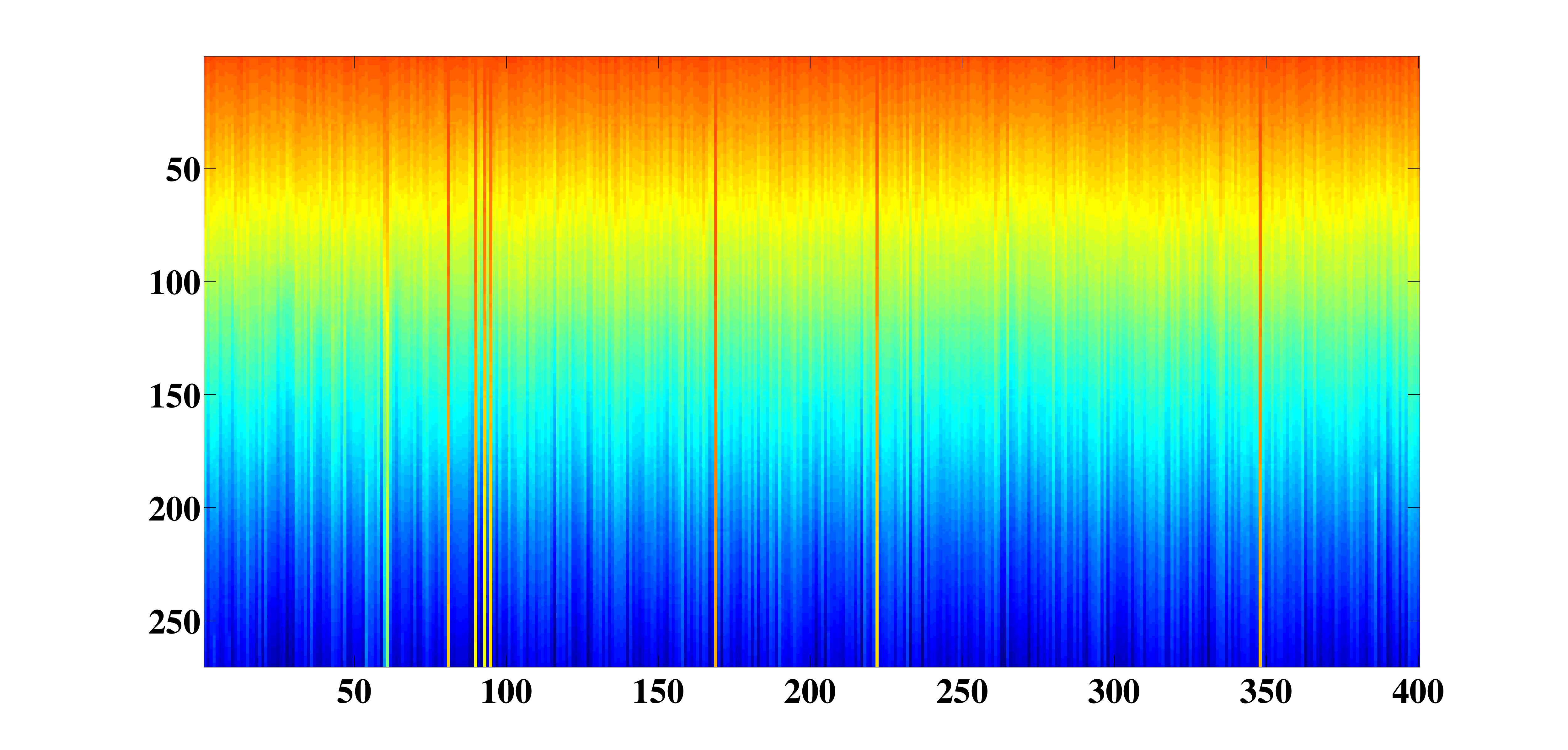}
            \caption[Network2]%
            {{\small Location 1, AP1}}    
        \end{subfigure}
        \hfill
        \begin{subfigure}[b]{0.2\textwidth}  
            \centering 
            \includegraphics[width=\textwidth]{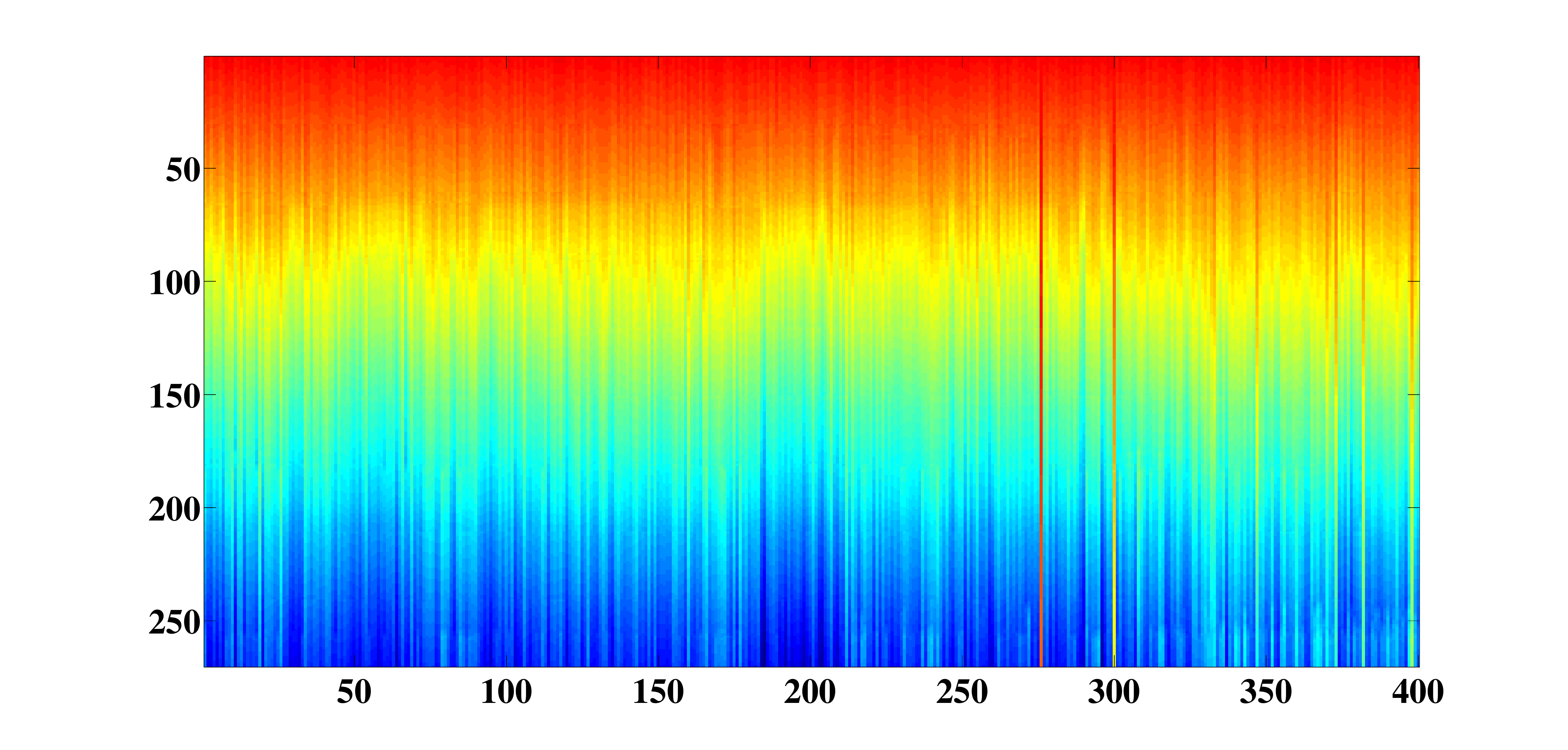}
            \caption[]%
            {{\small Location 1, AP2}}    
        \end{subfigure}
        \vskip\baselineskip
        \begin{subfigure}[b]{0.2\textwidth}   
            \centering 
            \includegraphics[width=\textwidth]{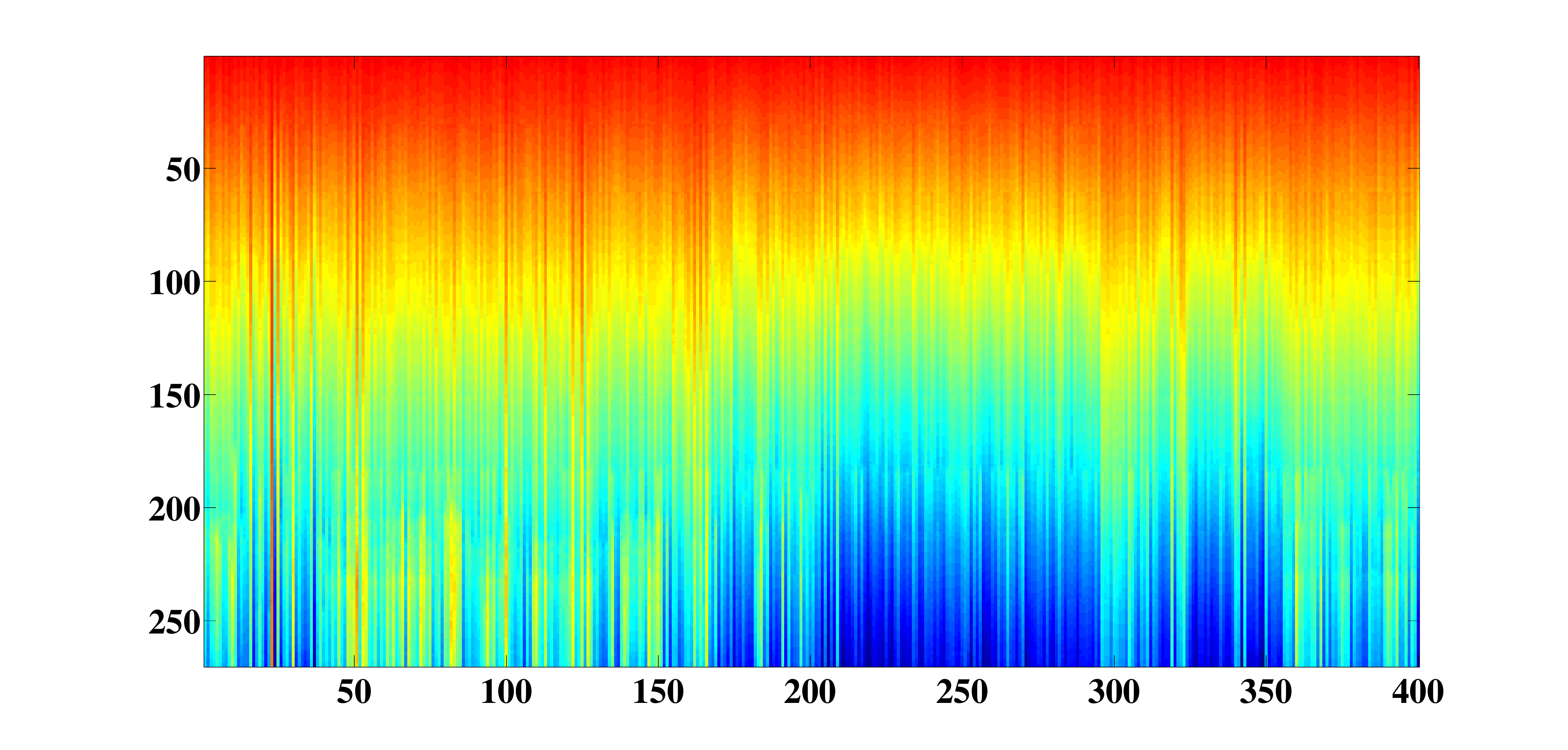}
            \caption[]%
            {{\small Location 2, Ap1}}    
        \end{subfigure}
        \hfill
        \begin{subfigure}[b]{0.2\textwidth}   
            \centering 
            \includegraphics[width=\textwidth]{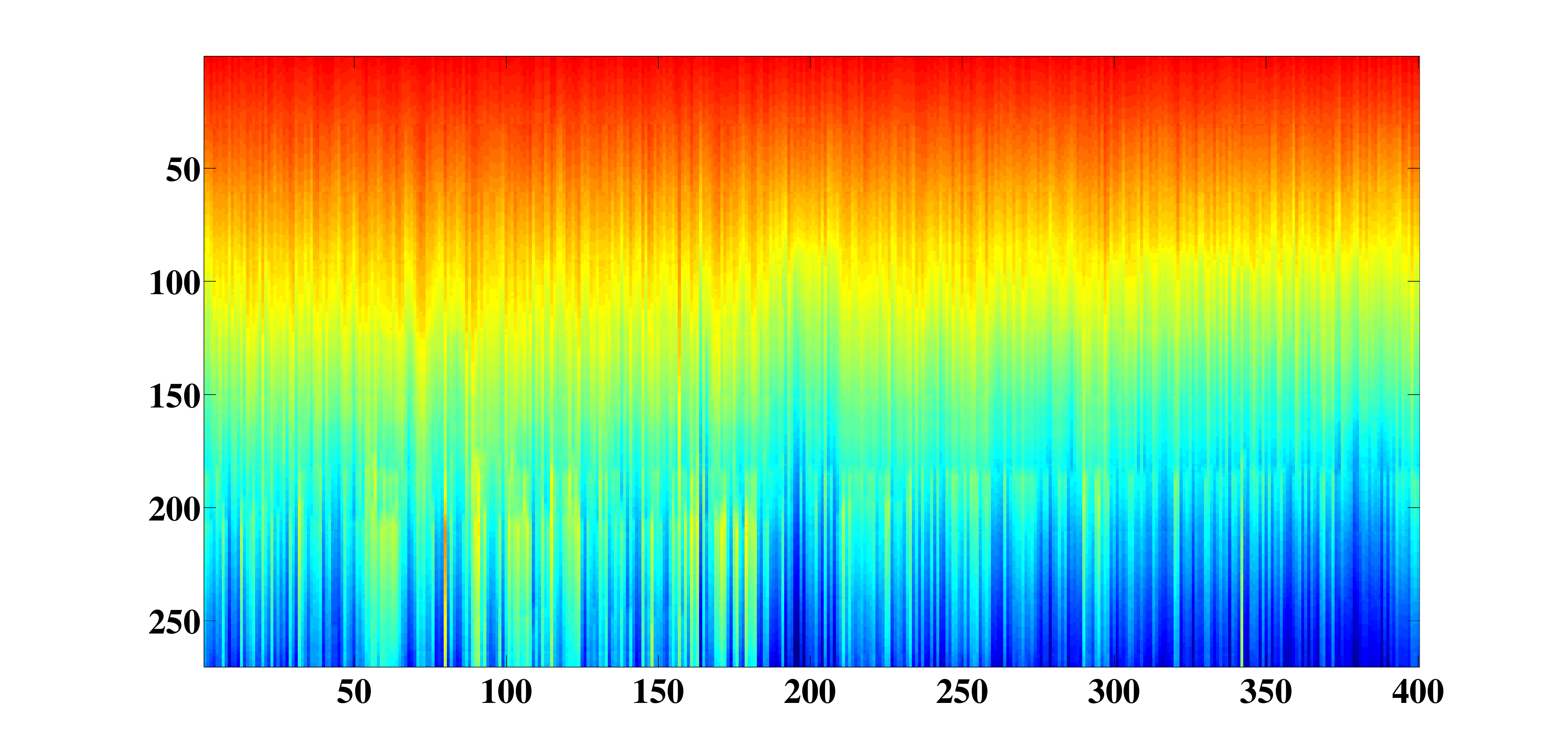}
            \caption[]%
            {{\small Location 2, AP2}}    
        \end{subfigure}
        \caption{Feature images of different locations using CSI phase.}
        \label{CSI_image_phase}
    \end{figure}
To validate this idea, we carry out some preliminary tests. We transform the CSI measurements from multiple OFDM channels of each AP-DP link into a feature matrix corresponding to an image. Since the OFDM channels are correlated, for each AP-DP link, CSI measurements received from multiple channels are also correlated. Therefore, instead of dealing the CSI measurements from each antenna channel independently, joint utilization of CSI measurements from all the antenna channels offers us to leverage channel correlations for improved localization. This motivates us to consider CSI measurements from multiple transmission-receiver antenna channels of each AP-DP link as features corresponding to the pixels of an image sample. We consider the CSI measurements from one transmission-receiver antenna channel as one RGB channel of an image. Therefore, from the two-dimensional perspective, each column corresponds to an image sample and for each sample, the CSIs in the rows corresponds to the pixel values of an image. Fig. \ref{CSI_image_amp} and \ref{CSI_image_phase} illustrates some CSI amplitude and phase feature images, respectively; obtained from multiple APs while a person is located at different locations. From the figures, it can be seen that the CSI amplitude and phase feature images from different locations have different patterns. In addition, for both the amplitude and phase, features corresponding to the same cell/location are different for two different APs. Therefore, CSI amplitude and phase feature images that are collected from different APs can be considered as good candidates for designing a device free localization system through multi-view learning approach, and thereby resulting in a better localization performance. 

\section{PRELIMINARIES}\label{sec:preli}

\subsection{Channel State Information}\label{sec:CSI}
In Multiple Input Multiple Output (MIMO) Orthogonal Frequency-Division Multiplexing (OFDM) technology, the narrow-band flat fading channel is modeled as, 
\begin{equation}
y = \boldsymbol{\mathscr{H}}x + \zeta,
\end{equation}
where $y$ and $x$ are the received and the transmitted signal vectors respectively, $\zeta$ is the noise vector and $\boldsymbol{\mathscr{H}}$ denotes the channel matrix. The channel matrix $\boldsymbol{\mathscr{H}}$ can be estimated by
\begin{equation}
\hat{\boldsymbol{H}} = \frac{y}{x},
\end{equation}
where $\hat{\boldsymbol{H}}$ represents the PHY layer CSIs over multiple sub-carriers. For one transmitter-receiver (Tx-RX) antenna pair, $\hat{\boldsymbol{H}}$ is a $S \times N$ matrix for each AP-DP link, where $S$ denotes the number of subcarriers for each antenna pair and $N$ is the number of measurements. CSI of a single subcarrier $k$ is a complex value \cite{CSi_phase}, 
\begin{equation}
h_{k}=R_{k}+j I_{k}=|h_{k}|e^{j sin\theta_{k}},
\end{equation}
where $R_{k}$ and $I_{k}$ are the in-phase and quadrature components, respectively; $|h_k|$ is the amplitude, and $\theta_k$ is the phase of $k$-th subcarrier. The amplitude response of subcarrier $k$ is  \(|h_{k}|=\sqrt{R_{k}^2 + I_{k}^2}\), and the phase response is computed by \(\angle h_{k}=\arctan(I_{k}/R_{k})\). We group CSIs of all Tx-Rx antenna pairs of each AP-DP link as,
\begin{equation}\label{H_all}
\boldsymbol{H}=[\hat{\boldsymbol{H}_{1}};\hat{\boldsymbol{H}_{2}};\ldots ;\hat{\boldsymbol{H}_{l}}],
\end{equation}
where $l$ is the index of Tx-Rx antenna pairs for each AP-DP link and $\boldsymbol{H}$ $\in$ $\mathbb{R}^{d \times N}$, where $d = S\times l$, the total number of subcarriers from all Tx-Rx antenna pairs. We can consider $\boldsymbol{H}$ as the feature image of CSI, where  each column corresponds to an image sample and for each sample, the CSIs in the rows corresponds to the pixel values of an image.

\subsection{CSI Phase Information}\label{subsec:phase}
CSI phase data extracted from the Intel 5300 NIC is highly random. The direct use of this phase data results in high error for indoor localization. This error stems from the hardware imperfection, specifically from the lack of synchronization of time and frequency of the transmitter and receiver. In order to overcome the error due to phase randomness, in this work we exploit the difference in phase values between two receiver antennas. For data packets that are received consecutively, this phase difference between two receiver antennas is highly stable in 5GHz 5300 NICs \cite{phaser_biloc19}. The measured CSI phase value $\theta_{k}$ from any subcarrier $k$ of one AP can be expressed as \cite{Precise_power, Speth},

\begin{equation}
\theta_{k} = \phi_{k} + k · (\lambda_{PB} + \lambda_{SF}) + \lambda_{CF},
\end{equation}
where $\phi_{k}$ is the original phase of subcarrier $k$ caused by the channel propagation, $i$ is the subcarrier index, $\lambda_{PB}$, $\lambda_{SF}$, and $\lambda_{CF}$ are phase errors resulted from the packet boundary detection (PBD), the sampling frequency offset (SFO), and central frequency offset (CFO), respectively. We aim to obtain the phase value $\phi_{k}$ by eliminating the impact of error parameters $\lambda_{PB}$, $\lambda_{SF}$, and $\lambda_{CF}$. 

Phase error $\lambda_{PB}$ is caused by the time shift $\tau_{PB}$ from the packet boundary detection uncertainty while the phase error $\lambda_{SF}$ is generated due to the offset of the sampling frequencies of the sender and the receiver. On the other hand, due to the hardware imperfection, the central frequency offset compensation is incomplete, which can cause CSI phase error $\lambda_{CF}$. Based on \cite{Speth}, it can be shown that,

\begin{equation}
\begin{aligned}
		& \lambda_{PB} = 2 \pi \frac{\Delta \tau}{N_f},\\
		& \lambda_{SF} = 2 \pi (\frac{T_{r}-T_{t}}{T_{t}})\frac{Ts}{Tu},\\
		& \lambda_{CF} = 2 \pi \Delta f T_{s} \eta,
	\end{aligned} \label{phase_errors}
\end{equation} 
where $N_f$ is the FFT size, $\Delta \tau$ is the packet boundary detection delay, $T_{r}$ and $T_{t}$ are the sampling periods of the receiver and the transmitter, respectively, $T_{u}$ is the data symbol length, $T_{s}$ is the total length of the guard interval and the data symbol, $\eta$ is the current packet sampling time offset, $\Delta f$ is the difference of center frequency between the transmitter and receiver. However, the value of $\Delta \tau$, $(\frac{T_{r}-T_{t}}{T_{t}})$, $\eta$, and $\Delta f$ in (\ref{phase_errors}) are unknown, since only physical layer CSI data are received from the off-the-shelf devices. Furthermore,  $\Delta \tau$ and $\eta$ are different for different packets, which causes variation in $\lambda_{PB}$, $\lambda_{SF}$, and $\lambda_{CF}$ over time. Hence, the original phase cannot be properly detected by the measured CSI phase. 

However, the difference in measured CSI phase values on a particular subcarrier between two receiver antennas in MIMO OFDM system is stable. This stability stems from the same clock and the same down-converter frequency of the receiver antennas of a particular Intel 5300 NIC device. For a particular subcarrier $k$, this in turn, results in the same central frequency difference, same delay in packet detection and same sampling period for the measured CSI phase. Hence, the difference in measured CSI phase between two antennas at subcarrier $k$,  can be approximated as,

\begin{equation}\label{ph_diff}
    \Delta \theta_{k} \approx \Delta \phi_{k},
\end{equation}
where $\Delta \phi_{k}$ is the phase difference of original phase between two adjacent antennas on subcarrier $k$. From (\ref{ph_diff}) it can be seen that the effect of random phase errors are minimized since the random terms $\Delta t$, $\eta$ and $\Delta f$ associated with $\lambda_{PB}$, $\lambda_{SF}$, and $\lambda_{CF}$ are eliminated. The phase differences are further shifted to be zero mean in order to ensure that initial phase offset errors for each packet are also minimized. Consequently, over different packets, $\Delta \theta_{k}$ becomes more stable compared to the individual CSI phase value. 

\subsection{Canonical Correlation Analysis }\label{subsec: CCA}
For multi-data processing, Canonical Correlation Analysis (CCA) is considered as one of the useful tools for finding a linear relationship between two feature sets \cite{hafiz_ref2}. CCA finds a common space for two views such that the correlation between these transformed feature sets are maximized in the common subspace.

Suppose that $n$ training feature vectors of the data from two different views are denoted by two matrices, $\boldsymbol{X}_1$ $\in$ $\mathbb{R}^{p \times n}$ and $\boldsymbol{X}_2$ $\in$ $\mathbb{R}^{q \times n}$, with dimension $p$ and $q$  for each training vector, respectively. For simplicity, we assume
that the observed samples are mean-centered. CCA aims to find a common subspace such that the pair-wise correlation across the two feature sets are maximized. In order to project the samples
from two views into the common subspace respectively, two linear transforms $\boldsymbol{w}_1$ and $\boldsymbol{w}_2$ are obtained by maximizing the correlation between $\boldsymbol{w}_1^T\boldsymbol{X}_1$ and $\boldsymbol{w}_2^T\boldsymbol{X}_2$ as below \cite{hafiz_ref2,hafiz_3}:
\begin{equation}\label{constraint_2}
	\begin{aligned}
		& {\underset{\boldsymbol{w}_1,\boldsymbol{w}_2}{\text{max}}} & & \boldsymbol{w}_1^{T}\boldsymbol{X}_1\boldsymbol{X}_2^T\boldsymbol{w}_2 \\
		& \text{subject to}	& & \boldsymbol{w}_1^{T}\boldsymbol{X}_1\boldsymbol{X}_1^T\boldsymbol{w}_1=1, \boldsymbol{w}_2^{T}\boldsymbol{X}_2\boldsymbol{X}_2^T\boldsymbol{w}_2=1.
	\end{aligned}
\end{equation}
Applying Lagrange multiplier on (\ref{constraint_2}), the optimization problem of CCA can be solved by a generalized eigenvalue problem as follows \cite{MCCA}:
\begin{equation}\label{matrix_eq1}
\begin{bmatrix}
     \boldsymbol{0}      & \boldsymbol{X}_1\boldsymbol{X}_2^T \\
   \boldsymbol{X}_2\boldsymbol{X}_1^T     &   \boldsymbol{0}   
\end{bmatrix}
\begin{bmatrix}
 \boldsymbol{w}_1\\
\boldsymbol{w}_2
\end{bmatrix}
= \lambda 
\begin{bmatrix}
    \boldsymbol{X}_1\boldsymbol{X}_1^T & \boldsymbol{0}\\
        \boldsymbol{0}   & \boldsymbol{X}_2\boldsymbol{X}_2^T\\
\end{bmatrix}
\begin{bmatrix}
  \boldsymbol{w}_1\\
\boldsymbol{w}_2
\end{bmatrix},
\end{equation}
where the degree of correlation between projections are reflected by the generalized eigenvalue $\lambda$. 

The CCA based approach described above is unsupervised. For pattern recognition problems, separating the classes is an important issue to consider. In CCA, the features are decorrelated, but the concept of class structure among the samples are not considered. In order to exploit class structures, discriminant CCA (DCCA) is proposed which takes into consideration both within-class and between-class correlation in CCA \cite{DCCA}. DCCA preserves the class structures for $C$ classes between two views through the following optimization problem:
\begin{equation}\label{dcca1}
	\begin{aligned}
		& {\underset{\boldsymbol{w}_1,\boldsymbol{w}_2}{\text{max}}} & & \boldsymbol{w}_1^{T}\boldsymbol{X}_1\boldsymbol{G}\boldsymbol{X}_2^T\boldsymbol{w}_2 \\
		& \text{subject to}	& & \boldsymbol{w}_1^{T}\boldsymbol{X}_1\boldsymbol{X}_1^T\boldsymbol{w}_1=1, \boldsymbol{w}_2^{T}\boldsymbol{X}_2\boldsymbol{X}_2^T\boldsymbol{w}_2=1.
	\end{aligned}
\end{equation}
where
\begin{equation}\label{dcca2}
\boldsymbol{G}= 
\begin{bmatrix}
   \boldsymbol{I}_{n_1 \times n1}  &  & & & \\
     & \ddots & & & \bigzero \\
      & &\boldsymbol{I}_{n_c \times n_c} &  & \\
     \bigzero& &  & \ddots &  \\
     & &  & & \boldsymbol{I}_{n_C \times n_C}
\end{bmatrix}.
\end{equation}
Applying Lagrange multiplier on (\ref{dcca1}), the optimization problem of DCCA can be solved by a generalized eigenvalue problem as follows \cite{DCCA}:
\begin{equation}\label{matrix_eq_DCCA}
\begin{bmatrix}
     \boldsymbol{0}      & \boldsymbol{X}_1\boldsymbol{G}\boldsymbol{X}_2^T \\
   \boldsymbol{X}_2\boldsymbol{G}\boldsymbol{X}_1^T     &   \boldsymbol{0}   
\end{bmatrix}
\begin{bmatrix}
 \boldsymbol{w}_1\\
\boldsymbol{w}_2
\end{bmatrix}
= \lambda 
\begin{bmatrix}
    \boldsymbol{X}_1\boldsymbol{X}_1^T & \boldsymbol{0}\\
        \boldsymbol{0}   & \boldsymbol{X}_2\boldsymbol{X}_2^T\\
\end{bmatrix}
\begin{bmatrix}
  \boldsymbol{w}_1\\
\boldsymbol{w}_2
\end{bmatrix}.
\end{equation}

\section{The MuDLoc System}\label{sec: proposed}
In MuDLoc the overall localization is performed through an offline phase and an online phase as described below.

\subsection{Offline Phase}
\subsubsection{Construction of CSI Amplitude and Phase Feature Image}\label{subsubsec: amp and ph feature image}

Exploiting bi-modal features of CSI, in terms of the amplitude and the phase from commodity WiFi device facilitates to utilize complete multipath features to achieve a high precision indoor localization system. In MuDLoc, the area is considered as a grid of small square cells. Let, there are \textit{C} cells and \textit{M} APs in that area of interest. In the offline stage, a set of CSI measurements are collected with the subject present in a cell, c , where c=1,2,...,C. Each cell can be considered as a class. Let class c has $n_c$ data samples. Next, CSI feature image $\boldsymbol{H}_{i}^c$ is generated for each cell using (\ref{H_all}), where i=1,2,...,M. $\boldsymbol{H}_{i}^c$ represents the effect of the presence of an entity on the $i$-th AP's CSI for the entity located at a particular position or cell, c. From $\boldsymbol{H}_{i}^c$, CSI amplitudes are extracted to generate amplitude feature image, $\boldsymbol{X}_i$ of size $d_{X_i} \times n$, where $n = \sum_{j=1}^{C} n_c $.
Similarly, the phase information is also extracted from $\boldsymbol{H}_{i}^c$ in order to generate CSI phase based feature image for each cell and then phase feature image, $\boldsymbol{Y}_i$ of size $d_{X_i} \times n$ is generated based upon the CSI phase difference of two adjacent receiver antennas for each AP using (\ref{ph_diff}).
These phase differences on a particular subcarrier between two receiver antennas in MIMO OFDM system are relatively stable compared to the raw phase information, since the effect of random phase errors are minimized as described in section \ref{subsec:phase}. 



Once both the amplitude and the phase difference-based feature images for all the APs in the area of interest are obtained, the system then exploits multi-view discriminant learning approach in order to obtain a discriminant common spaces for localization.

\subsubsection{Multi-view Discriminant Learning of CSI}
Good performance of the CCA-based indoor localization has been confirmed by the study on amplitude-based CSIs\cite{DisLoc}. However, only the CSI amplitude from a single AP (view) has been considered for location estimation. We consider that some common features should be shared by a set of CSI measurements recorded from multiple APs for a particular target cell. In particular, CSI data, extracted from multiple APs located at different locations reflect different characteristics of the patterns of amplitude and phase features affected by the presence of a subject at a particular cell. Therefore, such common features contained in the several real time CSI measurements could be more useful for location estimation of a test set in contrast to a single AP (view) based localization. 
The optimal amplitude and phase feature sets of CSIs are first learned from the joint spatial filtering of multiple sets of CSI amplitude and phase feature images, respectively; and are subsequently used in the feature fusion, where the transformed amplitude and phase feature images are stacked to obtain the complete feature set for cell recognition \cite{feature_fusion1}.

CCA based approach, described in section \ref{subsec: CCA} is only designed for two-view case, and thus the pairwise strategy is needed when applied to the multi-view scenario. Any generalization of the CCA to several sets has to be equivalent to the CCA in the case of 2 sets. MCCA is a generalization of CCA to more than two views of data, where the overall correlation among canonical variates from multiple sets of random variables is maximized through the optimization of the objective function of correlation matrix of the canonical variates \cite{MCCA,MCCA_ref1,KETTENRING}. The five most discussed  versions of MCCA  are: (1) SUMCOR, maximize the sum of all entries in the correlation
matrix; (2) MAXVAR, maximize the largest eigenvalue of the correlation matrix; (3) SSQCOR, maximize the sum of squares of all entries in the correlation matrix; (4) MINVAR, minimize the smallest eigenvalue of the correlation matrix; (5) GENVAR, minimize the determinant of the correlation matrix. Similar results are obtained for all of the five objective functions on a group dataset \cite{5_MvCCA, five_MvCCA}. This paper summarizes the classical sum of correlations generalization (SUMCOR) and MCCA is used as an abbreviation for SUMCOR maximization approach throughout the paper.
\cite{MCCA,MCCA_ref1}.

Let, for \textit{M} APs, multiple sets of random variables with n samples
of $d_i$ dimension are denoted by $\boldsymbol{X}_i$ $\in$ $\mathbb{R}^{{d_i} \times n}$, where, i = 1, 2, . . ., M. We assume that $\boldsymbol{X}_i$'s are  normalized to have zero mean and unit variance. MCCA aims to find a set of linear transforms $\boldsymbol{w}_i\big|_{i=1}^M$, to respectively
project the samples of M views \{$\boldsymbol{X}_1,. . . ,\boldsymbol{X}_M$\} to one common space, i.e., \{$\boldsymbol{w}_1^T\boldsymbol{X}_1,...,\boldsymbol{w}_M^T\boldsymbol{X}_M$\}. The total correlation in the common space is maximized as below:
\begin{equation}\label{MvCCA1}
	\begin{aligned}
		& {\underset {\boldsymbol{w}_1,...,\boldsymbol{w}_N}{\text{max}}} & & \rho={\underset{i \neq j}{\sum_{i,j=1}^{M}}}\boldsymbol{w}_i^T\boldsymbol{X}_i\boldsymbol{X}_j^T\boldsymbol{w}_j\\
		& \text{s.t.}	& &  \boldsymbol{w}_i^{T}\boldsymbol{X}_i\boldsymbol{X}_i^T\boldsymbol{w}_i=1, i=1,2,...,M \\
	\end{aligned}
\end{equation}
where $\boldsymbol{w}_i$ are the unknown transforms that have to be estimated for each matrix $\boldsymbol{X}_i$, which are \textit{M} known full-rank data matrices. The full-rank constraint of data matrices may be relaxed by regularizing the estimated covariance matrices \cite{full_rank}. We can rewrite Eq. (\ref{MvCCA1}) as,
\begin{equation}\label{MvCCA2}
	\begin{aligned}
			& {\underset {\boldsymbol{w}}{\text{max}}} & & \rho=\boldsymbol{w}^T(\boldsymbol{C-D})\boldsymbol{w}\\
		& \text{s.t.}	& &  \boldsymbol{w}^T\boldsymbol{D}\boldsymbol{w}=1,  \\
	\end{aligned}
\end{equation}
where,
\begin{equation}\label{C_eq}
 \boldsymbol{C}= 
\begin{bmatrix}
    \boldsymbol{X}_1\boldsymbol{X}_1^T  &  \ldots & \boldsymbol{X}_1\boldsymbol{X}_M^T \\
    \vdots & \ddots & \vdots \\
    \boldsymbol{X}_M\boldsymbol{X}_1^T & \ldots & \boldsymbol{X}_M\boldsymbol{X}_M^T
\end{bmatrix},
\end{equation}
\begin{equation}\label{D_eq}
\boldsymbol{D}= 
\begin{bmatrix}
    \boldsymbol{X}_1\boldsymbol{X}_1^T  &  \ldots & \boldsymbol{0} \\
    \vdots & \ddots & \vdots \\
    \boldsymbol{0} & \ldots & \boldsymbol{X}_M\boldsymbol{X}_M^T
\end{bmatrix},
\end{equation}
\begin{equation}\label{w_eq}
 \boldsymbol{w}=
\begin{bmatrix}
    \boldsymbol{w}_1 \\
    \vdots  \\
    \boldsymbol{w}_M
\end{bmatrix}.
\end{equation}
Applying  Lagrange  multiplier  on  (\ref{MvCCA2}),  the  optimization problem  of  MCCA  can  be  solved  by  a  generalized  eigenvalue problem. Using the Lagrange multiplier $\lambda$, the cost function $J$ is formed as below, and the unknown transforms $\boldsymbol{w}$ is found to maximize it:
\begin{equation}\label{MvCCA_sol1}
J= \boldsymbol{w}^T(\boldsymbol{C-D})\boldsymbol{w} + \lambda (\boldsymbol{w}^T\boldsymbol{D}\boldsymbol{w}-1).
\end{equation}
Taking the derivative with respect to $\boldsymbol{w}$, one can write,
\begin{equation}\label{MvCCA_sol}
(\boldsymbol{C-D})\boldsymbol{w} = \lambda \boldsymbol{D}\boldsymbol{w}.
\end{equation}
Eq. (\ref{MvCCA_sol}) represents a general eigenvalue decomposition problem and the largest eigenvector maximizes the cost function in Eq. (\ref{MvCCA_sol1}). Therefore, the eigenvector corresponding to the largest eigenvalue in the general eigen decomposition of Eq. (\ref{MvCCA_sol}) provides the solution for the  optimization problem  of  MCCA. The $\lambda$ in Eq. (\ref{MvCCA_sol}) can be obtained by left multiplication with $\boldsymbol{w^T}$, which, applying the constraint from Eq. (\ref{MvCCA2}), implies $\lambda=\rho$. Similar to CCA, the number of samples in each view for MCCA should be the same, and MCCA is also an unsupervised method.

In order to achieve discriminative common subspace for all views, GMA in \cite{GMA} proposed a general framework for multiview analysis, where the supervised structure of each view is preserved while keeping the projections of different views close to each other in the latent common space as follows:
\begin{equation}\label{GMA}
	\begin{aligned}
		& {\underset {\boldsymbol{w}_1,...,\boldsymbol{w}_N}{\text{max}}} & & \sum_{i=1}^{M}\alpha_i\boldsymbol{w}_i^{T}\boldsymbol{S}_i\boldsymbol{w}_i +{\underset{i \neq j}{\sum_{i,j=1}^{M}}}\beta_{i,j}\boldsymbol{w}_i^T\boldsymbol{X}_i\boldsymbol{X}_j^T\boldsymbol{w}_j\\
		& \text{s.t.}	& &  \sum_{i}^{M}\gamma_i\boldsymbol{w}_i^{T}\boldsymbol{X}_i\boldsymbol{X}_i^T\boldsymbol{w}_i=1,
	\end{aligned}
\end{equation}
where $\alpha_i$, $\beta_{i,j}$ and $\gamma_i$ are the balance parameters; and $\boldsymbol{S_i}$ is the between-class scatter matrix for the i-th view, which is defined as:
\begin{equation}
\boldsymbol{S_i}=\sum_{c=1}^{C}n_{c}^{i}(\mu_{c}^{i}-\mu^{i})(\mu_{c}^{i}-\mu^{i})^{T},
\end{equation}
where $\mu_{c}^{i}$ is the mean of class $c$ of i-th view, $\mu^{i}$ is the overall mean of all classes under i-th view, and $n_{c}^{i}$ are the samples of class $c$ for i-th view. In the objective function of (\ref{GMA}), the first part arises from the idea of classical Linear Discriminant Analysis (LDA) in order to exploit discriminant vectors in each view \cite{Fortunelda}. The positive term $\alpha$ is to bring a balance among the objectives, and hence usually set to 1 so that the joint objective will be unbiased towards optimizing $\boldsymbol{w}_i$. $\beta_{i,j} > 0$ is a tunable parameter to balance the relative significance between the
CCA part and the LDA part in (\ref{GMA}). Since all the constraints in (\ref{MvCCA1}) are nonlinear and there is no closed form solution in the current form, so the constraints are coupled with $\gamma =$ trace ratio, in order to obtain a relaxed version of the problem with a single constraint.

From (\ref{GMA}), it is seen that the class label information within each view are considered in GMA, which makes it discriminative for recognition across multiple views. However, only the discriminant information within each individual view are employed in GMA while the discriminant information from the inter-view are left unconsidered, which may degrade the performance of inter-view matching. As discussed in section \ref{subsec: CCA}, DCCA in \cite{DCCA} proposes an effective supervised feature extraction method for CCA, which exploits discriminant information between views. From (\ref{matrix_eq_DCCA}) it is seen that DCCA has a similar optimization objective like CCA. In order to effectively make full use of correlation information within each view and between different views, this work proposes to combine intra-view and inter-view discriminant correlation analysis, and therefore designs the following Generalized Inter-view and Intra-view Discriminat Correlation Analysis (GI$^{2}$DCA), which preserves both interview and intraview class structures as follows:
\begin{equation}\label{MvDCCA}
		\begin{aligned}
		& {\underset {\boldsymbol{w}_1,...,\boldsymbol{w}_N}{\text{max}}} & & \sum_{i}^{M}\alpha_i\boldsymbol{w}_i^{T}\boldsymbol{S}_i\boldsymbol{w}_i +{\underset{i \neq j}{\sum_{i,j=1}^{M}}}\beta_{i,j}\boldsymbol{w}_i^T\boldsymbol{X}_i\boldsymbol{G}\boldsymbol{X}_j^T\boldsymbol{w}_j\\
		& \text{s.t.}	& &  \sum_{i}^{M}\gamma_i\boldsymbol{w}_i^{T}\boldsymbol{X}_i\boldsymbol{X}_i^T\boldsymbol{w}_i=1,
	\end{aligned}
\end{equation}
where $\boldsymbol{G}$ is formulated according to (\ref{dcca2}). Applying  Lagrange  multiplier  on  (\ref{MvDCCA}),  the  optimization problem  of  GI$^{2}$DCA  can  be  solved  by  a  generalized  eigenvalue problem following the similar approach of MCCA:
\begin{equation}\label{MvDCCA_sol}
\boldsymbol{T}\boldsymbol{w} = \lambda \boldsymbol{\hat{D}}\boldsymbol{w},
\end{equation}
where $\boldsymbol{T}$ and $\boldsymbol{\hat{D}}$ are defined as follows, respectively:
\begin{equation}\label{T_eq}
 \boldsymbol{T}= 
\begin{bmatrix}
    \alpha_1\boldsymbol{S}_1  &  \ldots & \beta_{1,N}\boldsymbol{X}_1\boldsymbol{G}\boldsymbol{X}_N^T \\
    \vdots & \ddots & \vdots \\
    \beta_{N,1}\boldsymbol{X}_N\boldsymbol{G}\boldsymbol{X}_1^T & \ldots & \alpha_N\boldsymbol{S}_N
\end{bmatrix}.
\end{equation}
\begin{equation}\label{D_eq2}
\boldsymbol{\hat{D}}= 
\begin{bmatrix}
    \gamma_1\boldsymbol{X}_1\boldsymbol{X}_1^T  &  \ldots & \boldsymbol{0} \\
    \vdots & \ddots & \vdots \\
    \boldsymbol{0} & \ldots & \gamma_M\boldsymbol{X}_M\boldsymbol{X}_M^T
\end{bmatrix}.
\end{equation}
The number of non-negative eigenvalues for the general eigenvalue decomposition in Eq. (\ref{MvDCCA_sol}) is $r\leq min(d_1,. . .,d_M)$, with the assumption that all $\boldsymbol{X}_i$ are of full rank. It can be noted that, this GI$^{2}$DCA scheme requires finding the eigenvectors of matrices with $d_i \times d_i$ dimensionalities. Defining $d :=max_{i}d_{i}$, it can be checked that GI$^{2}$DCA incurs complexity of order $O(d^3M)$. Once the linear transforms $\boldsymbol{w}_i\big|_{i=1}^M$ are obtained through the above approach, the transformed features, $\boldsymbol{Z_i}$ are calculated by projecting $\boldsymbol{X}_i$ on the calculated $\boldsymbol{w}_i$ as,
\begin{equation}\label{transform_each}
\boldsymbol{Z}_i=\boldsymbol{w}_i^T\boldsymbol{X}_i, 
\end{equation}
where $\boldsymbol{Z}_i$ $\in$ $\mathbb{R}^{{r} \times n}$  and $i= 1,2,...,M$. Finally, the average canonical variate of the M datasets is calculated as,
\begin{equation}\label{general_Z}
\boldsymbol{Z} =\frac{1}{M}\sum_{i=1}^{M} \boldsymbol{Z}_i.
\end{equation}
$\boldsymbol{Z}$ in (\ref{general_Z}) represents the most common features that are shared among M sets of training data.

As described in section \ref{subsubsec: amp and ph feature image}, in the proposed MuDLoc system, the CSI amplitude and phase feature images collected from M APs (views) are denoted by $\boldsymbol{X_i}$ $\in$ $\mathbb{R}^{{d_{X_i}} \times n}$ and $\boldsymbol{Y_i}$ $\in$ $\mathbb{R}^{{d_{Y_i}} \times n}$, respectively. GI$^{2}$DCA is implemented to find multiple linear transforms (i.e. spatial filters) $\boldsymbol{w}_{\boldsymbol{X}_1}$,$\boldsymbol{w}_{\boldsymbol{X}_2}$, . . . ,$\boldsymbol{w}_{\boldsymbol{X}_N}$ that result in the maximization of overall correlation among the canonical variates $\boldsymbol{Z}_{\boldsymbol{X}_1}$,$\boldsymbol{Z}_{\boldsymbol{X}_2}$, . . . ,$\boldsymbol{Z}_{\boldsymbol{X}_N}$ obtained using (\ref{transform_each}). The transformed amplitude feature from M views, $\boldsymbol{Z_X}$ $\in$ $\mathbb{R}^{{r_X} \times n}$  is then calculated using (\ref{general_Z}). In a similar approach, the transformed phase feature from M views are calculated as $\boldsymbol{Z_Y}$ $\in$ $\mathbb{R}^{{r_Y} \times n}$. Finally, these bi-modal features extracted from multi-view CSI Data are stacked to obtain the single unified discriminant feature set, $\boldsymbol{Z}$ as follows: 
\begin{equation}\label{MvDF}
\boldsymbol{Z} = 
\begin{pmatrix} \boldsymbol{Z_X} \\ \boldsymbol{Z_Y} \end{pmatrix} .
\end{equation}
$\boldsymbol{Z}$ is called the Multi-view Discriminant Feature Image (MDFI) of CSI.  MDFI is more discriminative than any of the input feature image. GI$^{2}$DCA not only finds effective discriminant information over the multiple views of data but also eliminates redundant information within the features of each view, thereby improves the localization performance.

\subsection{Online Phase}
\begin{figure}[t]
\centering
\includegraphics[width=3.4 in]{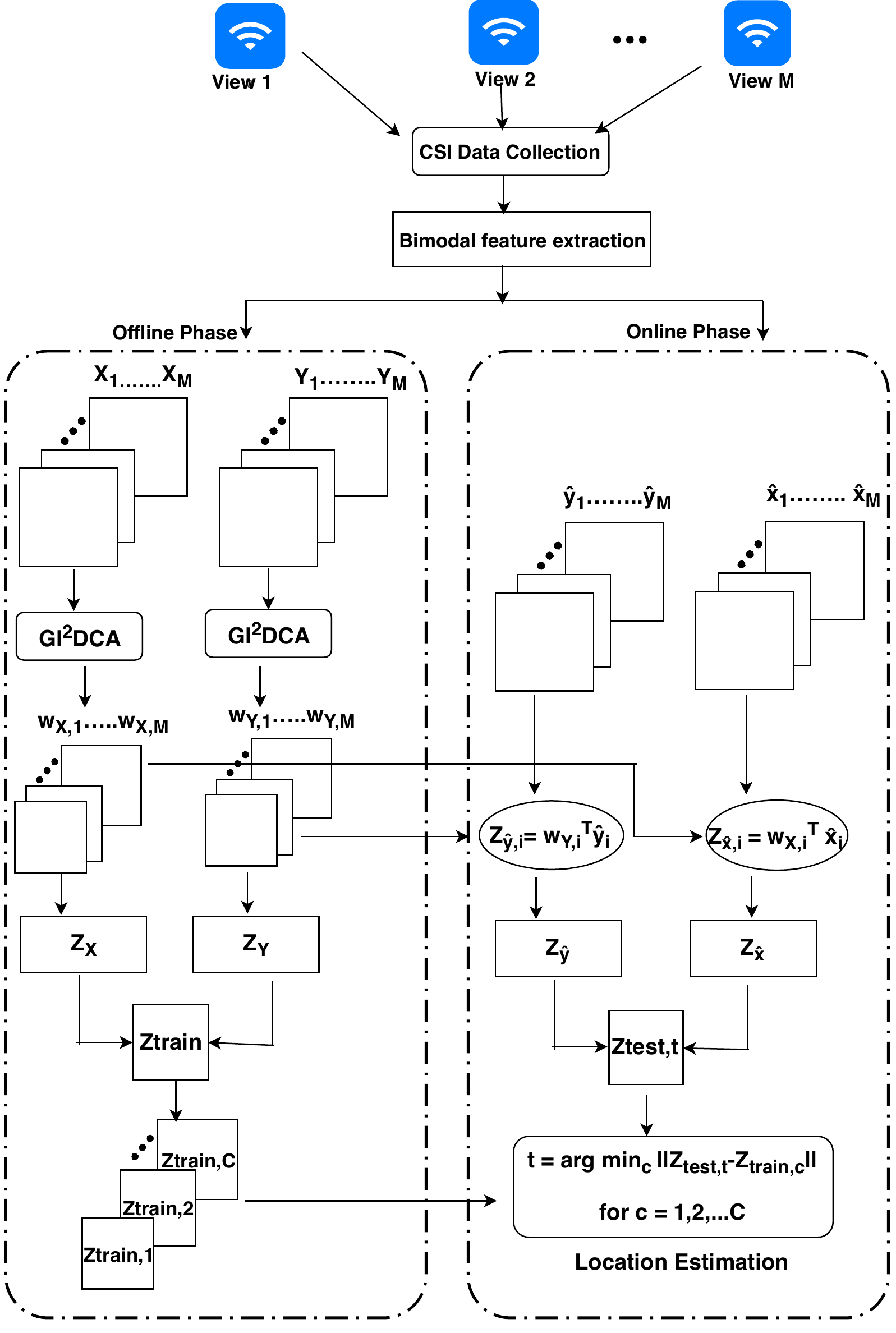}
\vspace{-0.2cm}
\caption{MuDLoc System Architecture.}
\vspace{-0.4cm}
\label{System_Architecture}
\end{figure}

After the aforementioned calibration procedure of training feature optimization in the offline phase, the linear transformations for the amplitude feature, $\hat{\boldsymbol{x}}_i$ of test location $t$ are calculated as,
\begin{equation}\label{x}
\boldsymbol{Z_{\hat{\boldsymbol{x}}_i}} = \boldsymbol{w_{X}}_i^T\hat{\boldsymbol{x}}_i,i=1,2,....,M.
\end{equation}
Simialrly, the linear transformations for the phase feature, $\hat{x}_i$ of test location, $t$ are calculated as, 
\begin{equation}\label{y}
\boldsymbol{Z_{\hat{y}_i}} = \boldsymbol{w_{Y_i}}^T\hat{\boldsymbol{y}}_i.
\end{equation}
The average canonical variate, $\boldsymbol{Z_{\hat{\boldsymbol{x}}_i}}$ and $\boldsymbol{Z_{\hat{\boldsymbol{y}}_i}}$ for test location $t$ are then calculated using (\ref{general_Z}). Finally the MDFI for test location $t$ is calculated as,
\begin{equation}\label{MvDF_test}
\boldsymbol{Z}_{test, t} = 
\begin{pmatrix} \boldsymbol{Z_{\hat{x}}} \\ \boldsymbol{Z_{\hat{y}}} \end{pmatrix} 
\end{equation}
Finally, the test location's cell Id is then recognized using the simple but efficient Euclidean Distance (ED) based similarity measure as following,
\begin{equation}
\argmin_c \|\boldsymbol{Z}_{test, t}-\boldsymbol{Z}_{train, c}\|_2,
\end{equation}
where, $c \in [1,2,...,C]$. The overall system architechture for MuDLoc system is shown in Fig. \ref{System_Architecture}.

\section{Experimental Study}\label{sec:Experimental Study}
\subsection{Experimental Configuration}
The MuDLoc system is implemented with Intel 5300 commodity Wi-Fi device and extensive experiments are conducted to valid its effectiveness. In order to measure CSI data in 5 GHz band, the system uses Lenovo laptops as the access points (AP) and a desktop computer as the detection point (DP) or mobile device. Both devices are equipped with an Intel 5300 Network Interface Card (NIC). The operating system is Ubuntu desktop 14.04 LTS OS. The access points are set in the monitor mode. The mobile device is set in the injection mode and uses one antenna to transmit
data. CSI data from multiple APs are obtained by using the packet injection technique
based on LORCON version 1. A host PC (Intel i7-4790CPU 3.60 GHz, 8GB RAM) serves as the centralized server for location estimation. Using the Linux 802.11n tool \cite{Halperin,Halperin1}, for each AP, the DP collects CSI data for 30 subcarriers for each Tx-Rx antenna pair. Therefore, $3 \times 3$ Tx-Rx antenna pairs for each AP-DP link is utilized. Finally, the amplitude and phase data are extracted for the training and test stages as described in Section \ref{subsubsec: amp and ph feature image} in order to implement MuDLoc     with GI$^{2}$DCA approach.  

The performance of MuDLoc is verified in various scenarios
and the resulting location errors in different environments are compared with several benchmark schemes. It is found from the experimental results that in an
open indoor space, where there are fewer or no obstacles in the area of interest, the performance of indoor localization is better than that in
a complex environment where there are fewer LOS paths. The experimental results are presented from two typical indoor
localization environments, as described in the following.

\begin{figure}[t]
\centering
\includegraphics[width=2.7in]{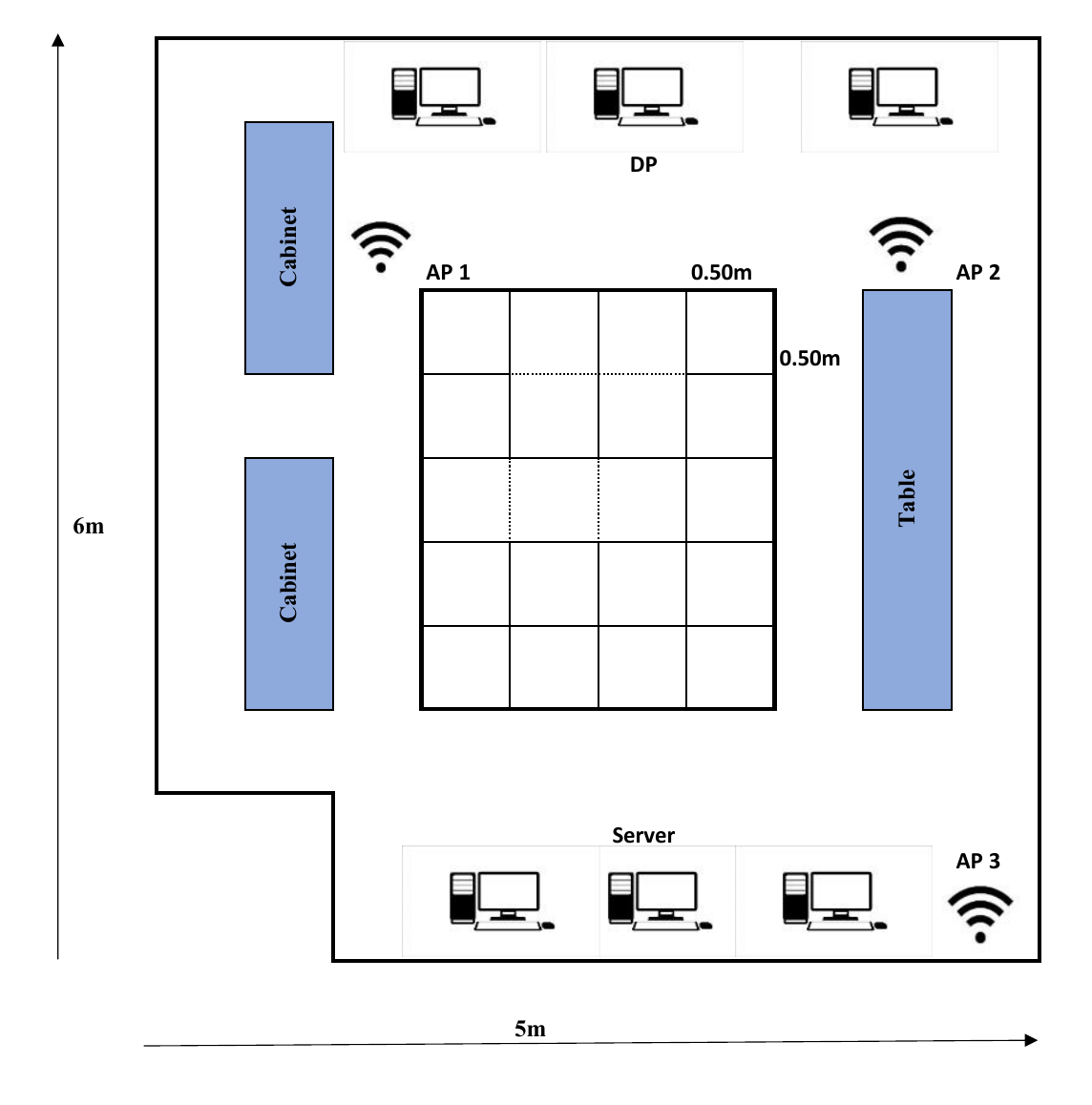}
\caption{The Layout of the Testbed in a Research Laboratory.}
\label{floor1}
\end{figure}

1) Research Laboratory: This is a research laboratory with an area of $6 m$ $\times$ $5 m$ in the CoRE Building of Rutgers University. Fig. \ref{floor1} shows the testbed layout of MuDLoc in the research laboratory. The lab is a cluttered environment, equipped with typical office facilities like desks, shelfs, desktops, chairs etc., which block most of the LOS paths and form a complex radio propagation environment. The area is virtually partitioned into 20 uniform square grids/cells, each of which is $0.50m$ $\times$ $0.50m$ in size. Total 20 training and 20 testing locations are considered for experiments. 5 different APs are placed in 5 different random places in order to exploit the multi-view approach.

\begin{figure}[t]
\centering
\includegraphics[width=2.9in]{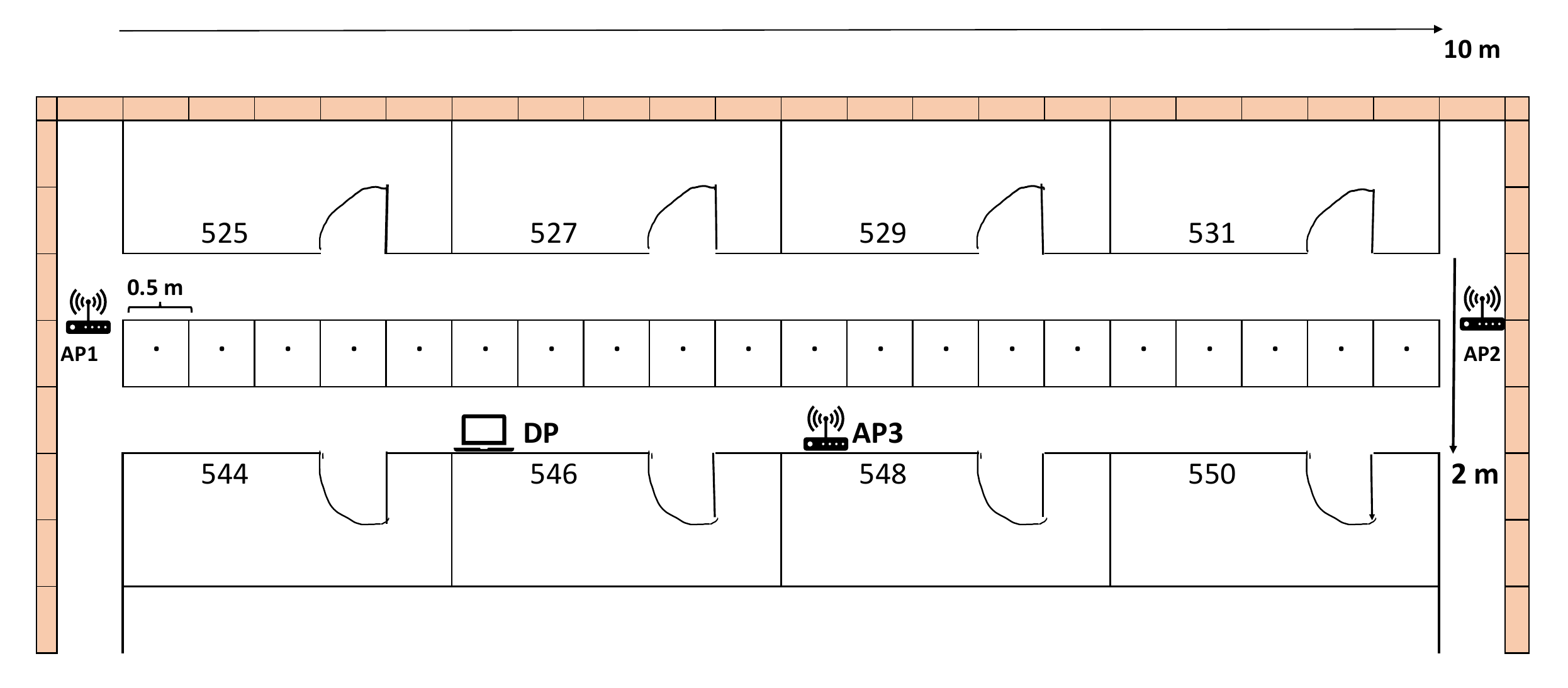}
\caption{The Layout of the Testbed in a Corridor.}
\vspace{0.7cm}
\label{corridor}
\end{figure}

2) Corridor: This is a long corridor at fifth floor of CoRE Building of Rutgers University with dimension $2 m$ $\times$ $10 m$. The corridor we choose is almost empty; therefore, most of the measured locations have LOS receptions. As in Fig. \ref{corridor}, we place 3 APs at different random locations on the floor to measure CSI data. $20$ positions are chosen uniformly scattered with half-meter spacing along a straight line for the corridor experiments. 

All experiments are conducted during weekdays. The CSI packets are received at 1s interval and we record for $5$ minutes for each cell. We take $300$ packet samples for each cell position. To take in to account the time domain variation, we conduct $10$ independent measurements on $10$ different days and compute the mean value for performance evaluation. For each location, we get 3000 samples for each of the APs (views). These time domain samples collected for each of the APs are grouped in order to create feature images for each location. The entire dataset is partitioned into training sets, validation sets and test sets using a ratio of 6:2:2. Following the suggestions in \cite{GMA}, $\mu$ is set to 1, $\gamma$ is set as trace ratio, and 5-fold cross validation are used to select tuning parameter $\beta$ among [0, 1000]. Four representative schemes are built from the literature, i.e., PC-DfL \cite{Xu_mob_com}, Pilot \cite{Pilot}, Pairwise CCA (PWCCA) \cite{hafiz_ref2} and MCCA \cite{MCCA}, which are
discussed in Section \ref{sec:introduction}. In order to ensure a fair comparison, same dataset obtained for the 5 GHz band is used by all the schemes. Extensive experiments with
the schemes are conducted in the above two representative indoor
environments to evaluate the performance of the proposed method.

\begin{table}[t]
  \begin{center}
    \caption{Comparison of mean distance error and standard deviation for different schemes in the laboratory environment}
    \label{tab:accuracy_lab}
    \vspace{0.3cm}
    \begin{tabular}{c|c|c}
      \toprule 
      \textbf{Algorithms} & \textbf{Mean distance error (m)} & \textbf{Std. dev. (m)} \\
      \midrule 
      PC-DfL & 1.3000 & 0.8346 \\
      Pilot & 1.01667 & 0.8242 \\
      PWCCA & 0.8665 & 0.7922 \\
      MCCA & 0.7032 & 0.4919 \\
      MuDLoc (GI$^{2}$DCA) & 0.2449 & 0.4449 \\
     \bottomrule 
    \end{tabular}\vspace{0.4cm}
  \end{center}
\end{table}
\subsection{Localization Performance}
\label{sec:results}
First the performance of MuDLoc system  with proposed GI$^{2}$DCA approach is evaluated in terms of mean distance error and standard deviation; and are compared with the RSS-based localization approach PC-DfL \cite{Xu_mob_com} and CSI-based approach Pilot \cite{Pilot}. The proposed system is also compared with related existing multi-data processing methods including pairwise CCA (PWCCA) \cite{hafiz_ref2} and MCCA \cite{MCCA}, when applied for indoor localization. Among them, PWCCA, is a two-view method; therefore, the pairwise strategy for multi-view classification is exploited for comparison. The results are presented for laboratory and corridor scenarios in Table \ref{tab:accuracy_lab} and \ref{tab:accuracy_corridor}, respectively. For the laboratory scenario, where there exists abundant multipath and shadowing effect, the mean error of MuDLoc is 0.2449 m and the STD error is 0.4449 m, as shown in Table \ref{tab:accuracy_lab}. In the corridor environment, where there exists more LOS receptions, the mean error of MuDLoc is 0.15 m and the STD error is 0.3095 m, as shown  in Table \ref{tab:accuracy_corridor}. MuDLoc outperforms the other multi-view learning approach, PWCCA and MCCA in both scenarios. MuDLoc achieves a $65\%$ improvement over MCCA by exploiting a inter-view and intra-view discriminnat learning approach in multi-view analysis. Moreover, all the CSI based schemes outperforms the RSS based approach, i.e., PC-DfL. The latter has a mean error of 1.3 m in the laboratory scenario and 1.16 m in the corridor scenario.


\begin{table}[t]
  \begin{center}
    \caption{Comparison of mean distance error and standard deviation for different schemes in the corridor environment}
    \label{tab:accuracy_corridor}
    \vspace{0.3cm}
    \begin{tabular}{c|c|c}
      \toprule 
      \textbf{Algorithms} & \textbf{Mean distance error (m)} & \textbf{Std. dev. (m)} \\
      \midrule 
      PC-DfL & 1.1589 & 0.6469  \\
      Pilot & 1.1349 & 0.6581 \\
      PWCCA & 0.7192& 0.7718 \\ 
      MCCA & 0.6888 & 0.6040 \\
      MuDLoc(GI$^{2}$DCA) & 0.1500 & 0.3095 \\
     \bottomrule 
    \end{tabular}\vspace{-0.4cm}
  \end{center}
\end{table}

\begin{figure}[t]
\centering
\includegraphics[width=3in]{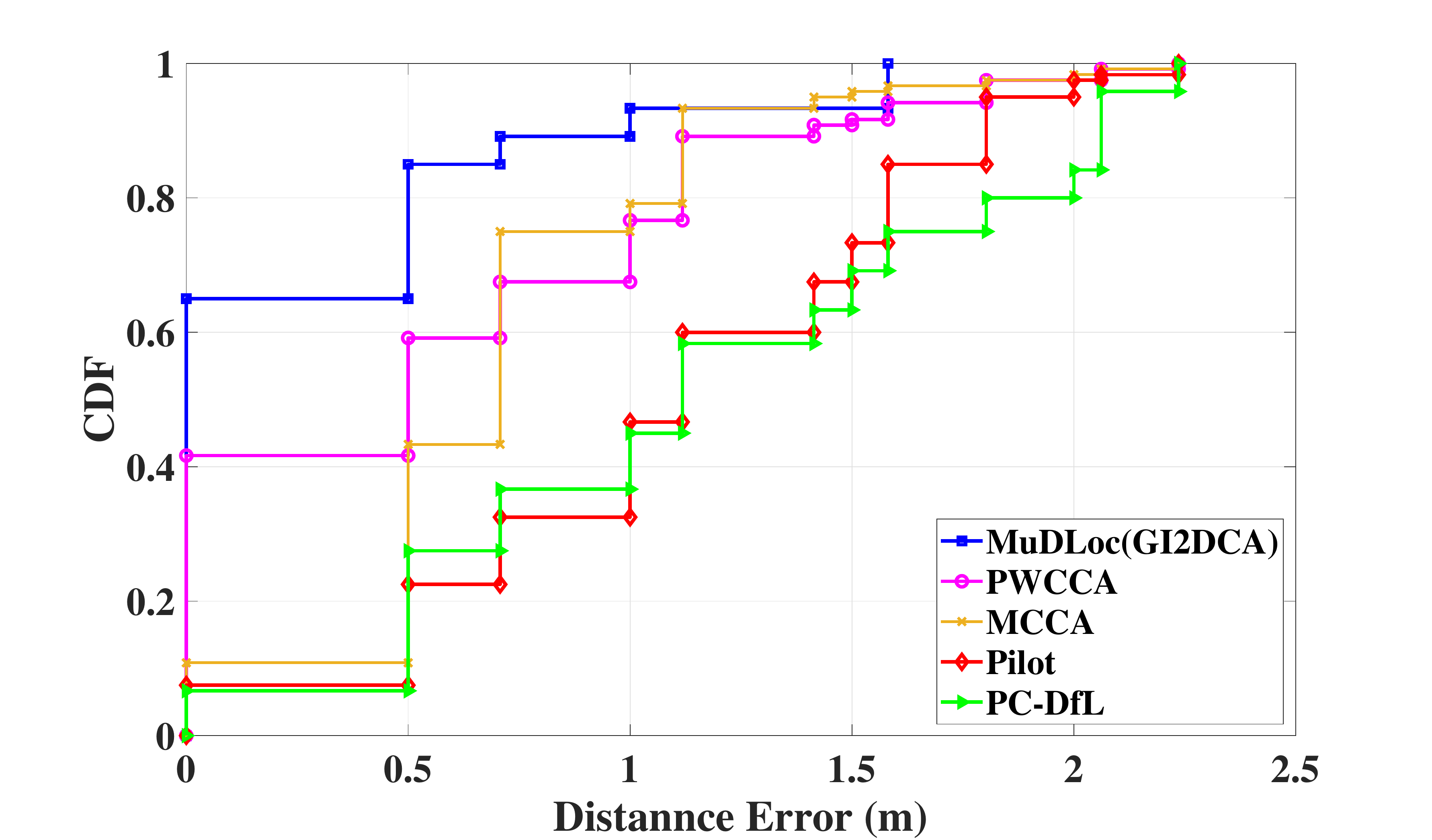}
\caption{CDF of Distance Error for Laboratory Scenario}
\label{cdfplot}
\end{figure}

Fig. \ref{cdfplot} presents the CDF of distance errors for different methods in the laboratory environment . MuDLoc has 65\% of the test locations having an error less than or equal to 0.5 m, while that for the other methods is 42\% or less. We also find that approximately 90\% of the test locations for MuDLoc have an error under 1 m, while the percentage of test locations having a smaller error than 1 m are 75\%, 68\%, 37\% and 33\% for MCCA, PWCCA, Pilot and PC-DfL, respectively. Thus, MuDLoc achieves the best performance in terms of distance error in this experiment. 

\begin{figure}[t]
\centering
\includegraphics[width=3in]{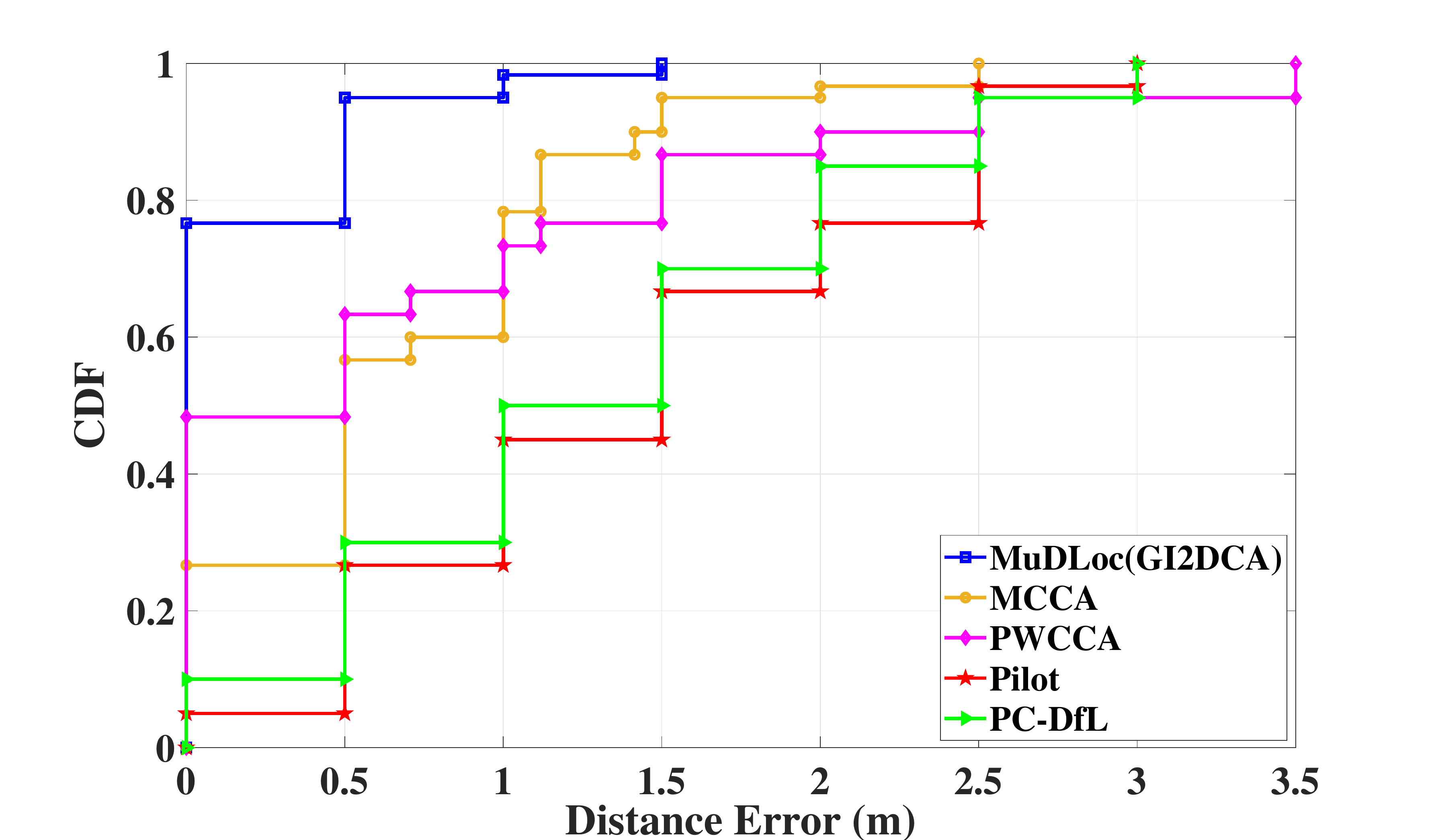}
\caption{CDF of Distance Error for Corridor Scenario}
\label{cdfplot_corr}
\end{figure}

In Fig. \ref{cdfplot_corr}, the CDF of distance errors for different methods in the corridor environment are presented. With MuDLoc, 78\% of the test positions have an error smaller than 0.5m m, while
with MCCA, PWCCA, Pilot and PC-DfL, close to 27\%, 48\%, 5\%  and 10\% of the test positions, respectively, have an error smaller than 0.5 m. Results also show that approximately 95\% of the test locations for MuDLoc have an error under 1 m, while that for the other methods is 67\% or less. Thus, MuDLoc also achieves the best performance for corridor environment. This is because the other methods are either designed to work with single AP or consider the average value for multiple APs. Morover,  all other methods use only the amplitude feature of CSI or RSS value for localization, while MuDLoc exploits CSIs from multiple APs through multiview discriminant analysis and fuses transformed amplitude and phase-based features of CSI into a single feature, which is more discriminative than the individual ones. This feature fusion method reduces the redundant information between two input features, and therefore will be more effective for better localization.


\begin{figure}[t]
\centering
\includegraphics[width=3in]{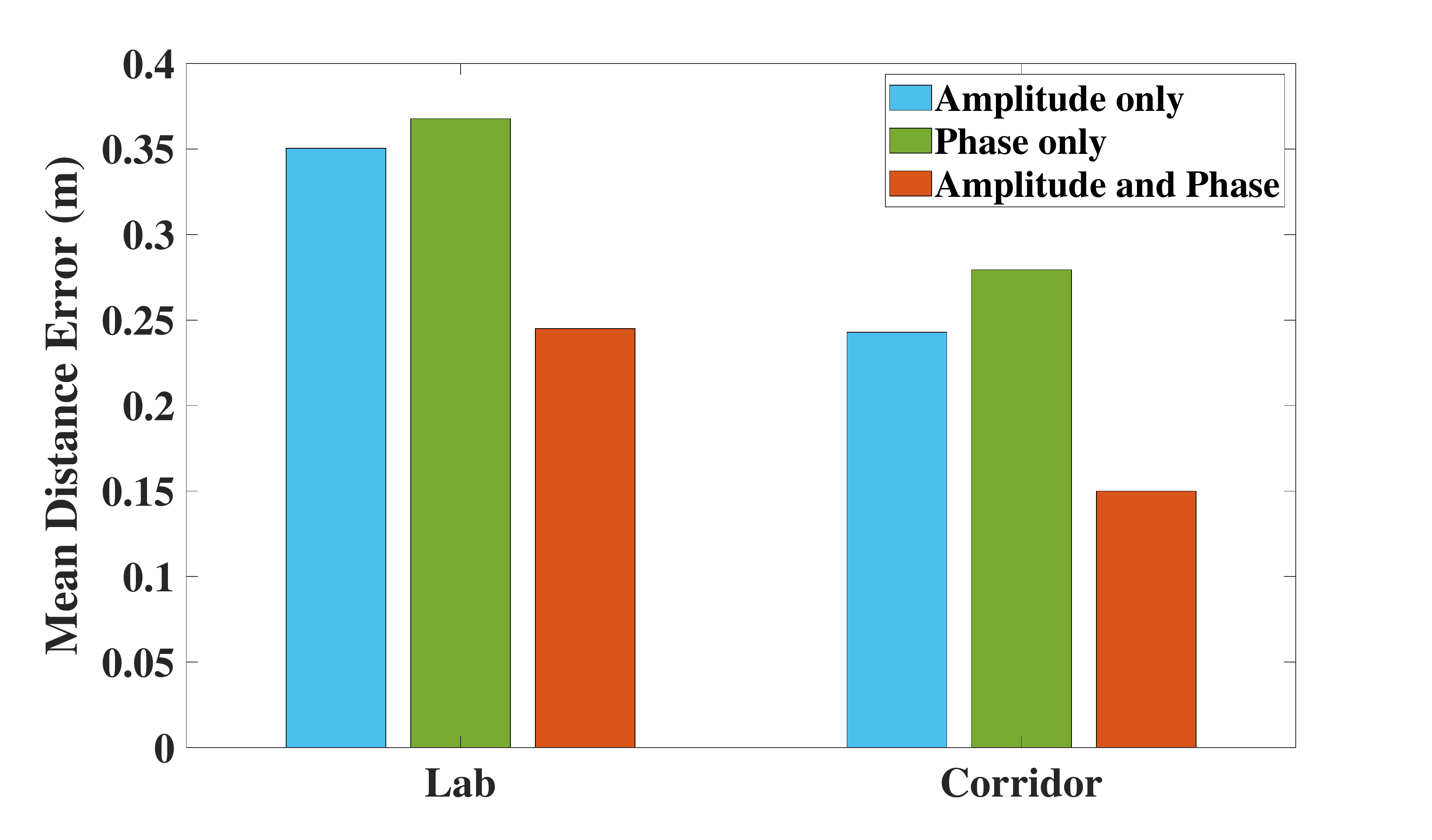}
\caption{Mean Distance Error for different modality of CSI measurements}
\label{amp_ph_both}
\end{figure}

\subsection{Impact of Using Bi-modal Features of CSI}
The proposed MuDLoc system utilizes CSI as observation measurements for indoor localization, which provides amplitude and phase information. The contribution of these bi-modal features of CSI data are analyzed through the evaluation of the system performance when using amplitude information, phase information, and both the amplitude and phase information of CSI. Fig. \ref{amp_ph_both} reveals that The performance for corridor scenario is always better than the laboratory scenario in all the cases due to more LOS reception for most of the measured positions. However, the MuDLoc system with GI$^{2}$DCA approach could achieve reasonable localization accuracy for both indoor scenarios even using only one type of measurement (either amplitude or phase). The mean distance error can be further decreased to as low as 0.25 m for laboratory scenario and 0.15 m for corridor scenario if using both the amplitude and phase measurements. Thus, utilizing the bimodal features of CSI in terms of amplitude and phase information from MIMO OFDM system facilitates to improve the localization performance to a great extent. 

\subsection{Impact of The Number of Samples}
The effect of the number of packets used in the online test phase of MuDLoc are also evaluated in the study. In the experiments, the location of the target is estimated using different numbers of
packets for the two indoor environments. The experiments are performed using 300, 400, 500, and 600 packets in the online test for location estimation. In Table \ref{tab:samples}, the mean distance errors for different numbers of packets in the laboratory and corridor experiments are shown. Results show that the mean distance error in the corridor experiment is lower than that in the laboratory experiment for different amounts of packets. Moreover, with the increase of packets,
the mean distance error for both experiments is decreased. It can be noted that the maximum distance errors for the
laboratory and corridor experiments are 0.2449 m and 0.1500 m, respectively, while the minimum distance errors for the laboratory and corridor experiments are 0.1893 m and 0.1002 m, respectively. In fact, for both experiments, with the increase in packets, the decrease of mean distance error is small. Therefore, 300 packets were chosen for the test phase in the proposed system for both indoor environments, with which the system can obtain a satisfactory localization performance with a lower computational complexity.

\begin{table}[t]
  \begin{center}
    \caption{Mean distance error versus the number of packets used in online test phase }
    \label{tab:samples}
    \vspace{0.3cm}
    \begin{tabular}{l|p{1.2cm}|p{1.2cm}|p{1.2cm}|p{1.2cm}}
      \toprule 
      \textbf{\# of packets} & \textbf{300} & \textbf{400} & \textbf{500} &  \textbf{600} \\
      \midrule 
     Laboratory & 0.2449 m & 0.2371 m & 0.2053 m & 0.1893 m \\
      Corridor & 0.1500 m & 0.1288 m & 0.1093 m & 0.1002 m\\
      \bottomrule 
    \end{tabular}\vspace{-0.4cm}
  \end{center}
\end{table}

\subsection{Impact of The Number of Views (APs)}
\begin{figure}[t]
\centering
\includegraphics[width=3in]{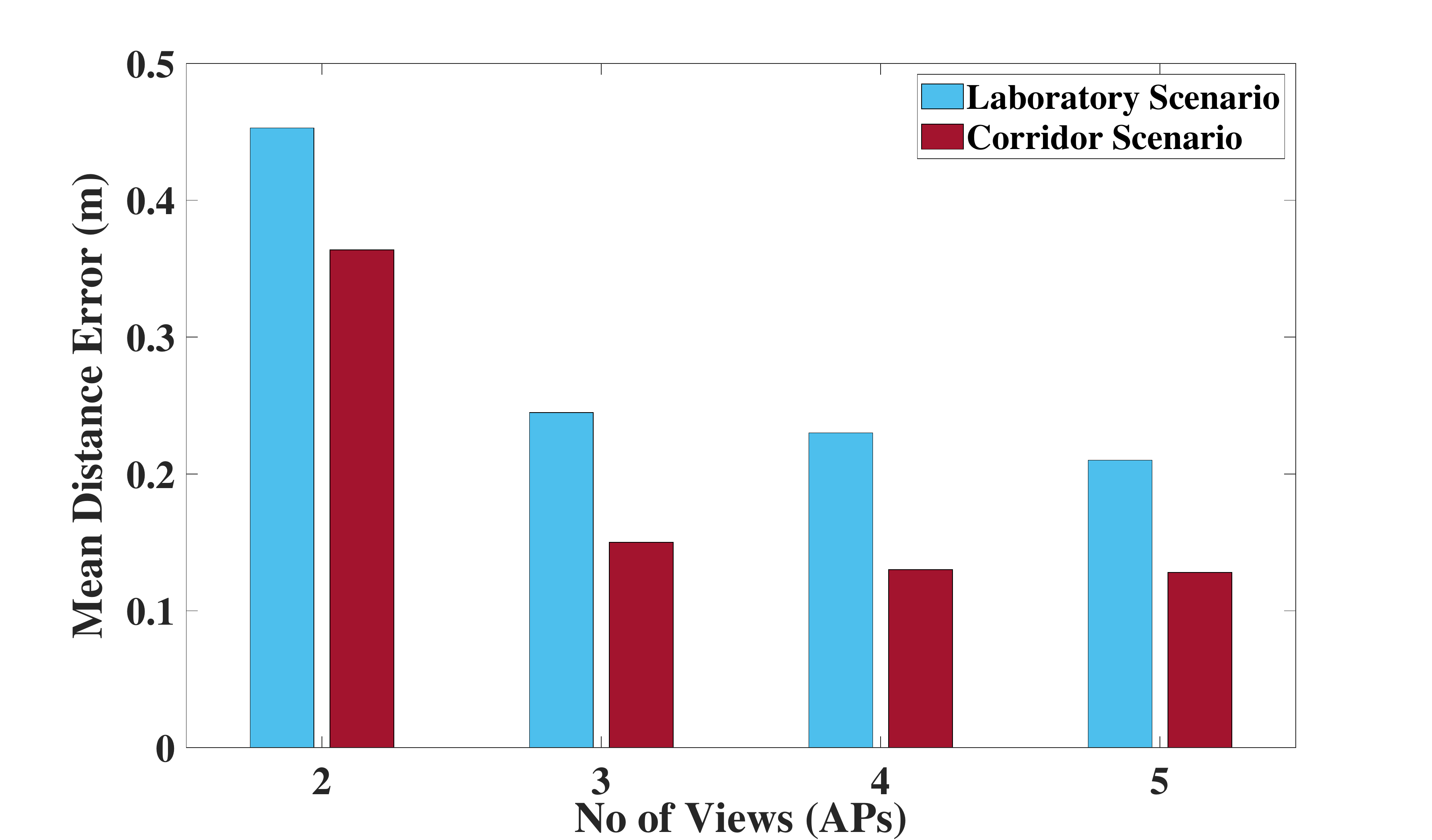}
\caption{Mean distance error in laboratory for different no of views (APS)}
\label{APs_lab}
\end{figure}

Finally, the impact of the number of views (APs) on localization performance is evaluated for the proposed method in the two indoor environments. The experiments are conducted with multiple views (APs) to evaluate the performance of the proposed system. Fig. \ref{APs_lab} presents the performance of MuDLoc (GI$^{2}$DCA) system in terms of the mean distance errors for different number of APs in laboratory and corridor environments. It is noticed that, with the increase in number of views/APs, the mean error is decreased for both indoor deployments. This is because, the more the number of APS, the richer the multipath information that can be obtained to estimate the location. However, results show that, for both environments, the decrease in mean distance error are relatively small when the number of AP is increased to 4 or more. Since MuDLoc can obtain fairly low localization errors using only 3 APs for both indoor deployments, this work considers using 3 AP for exploiting the multi-view approach in order to achieve the higher localization accuracy with lower deployment cost.

Although the MuDLoc scheme with three or more AP achieves lower
mean distance errors, it takes more time for processing the CSI
values from multiple AP as input data for each packet. We evaluate the average processing time to estimate the target position in the test phase using 300 received packets. The processing time is measured as the CPU occupation time for the MATLAB R2016a program running on the
centralized server. Results in Table \ref{tab:AP_time} show that, in the laboratory scenario, the single-view scheme (i.e. Pilot) takes 1.89 s, on average, to estimate the target position, whereas the multi-view scheme (MuDLoc) takes around 2.18 s, 2.35 s, 2.73 s and 3.09 s for processing CSI values from 2, 3, 4, and 5 APs, respectively, as input data to estimate the location. Therefore, with the increase in no of views (APs), the execution time increases. As shown in Fig. \ref{APs_lab}, MuDLoc can obtain fairly low localization errors using only 3 APs for laboratory scenario and the mean processing time is 2.35 s, which is lower than that for 4 AP or 5 AP systems. Moreover, the difference in execution time is small when compared with single view approach, although the multi-view approach processes three times input data than that in the single view (single-AP) scheme. The three-view MuDLoc takes about 29\%
extra processing time than single-view approach, but it can achieve a 83\% improvement in
localization precision for laboratory environment and the latter is generally more important for indoor localization.

\section{Conclusions}
\label{sec:Conclu}
This paper presents MuDLoc, a multi-view discriminant learning approach for indoor localization that exploits both the amplitude and the phase information of CSI in MIMO OFDM systems. In MuDLoc, CSI information for all the subcarriers from $3 \times 3$ MIMO channels are collected from multiple APs and analyzed with a multi-view learning approach to extract joint spatial features. In order to take discriminant information across multiple cells/locations and multiple views into account, the proposed MuDLoc system implements GI$^{2}$DCA, which preserves both inter-view and intra-view class structures through a discriminant correlation analysis. Both amplitude and phase information are utilized in order to exploit the complete multipath information from CSI measurements. These discriminant features extracted from the GI$^{2}$DCA are used for effective, high accuracy device free indoor localization by transforming the localization problem into a cell classification problem using pattern matching. The proposed MuDLoc scheme was validated in two representative indoor environments and was found to outperform several existing RSS and CSI based  localization schemes. The effect of different modalities of CSI data as well as system parameters on MuDLoc performance are examined. It was found that MuDLoc can achieve good performance with lower localization error under such scenarios.

\begin{table}[t]
\center \caption{Comparison of mean processing time for single-view vs, multi-view scheme in the laboratory environment}
\label{tab:AP_time}
    \vspace{0.3cm}
\def\arraystretch{1.5}
\begin{tabular}{|c|c|c|c|c|c|}
\hline
\multirow{2}{*}{No. of views} & \multicolumn{1}{c|}{single-view} & %
    \multicolumn{4}{c|}{multi-view (MuDLoc)} \\
\cline{2-6}
& Pilot & 2 AP & 3 AP & 4 AP & 5 AP \\
\hline
Proc. time (s) & 1.89 & 2.18 & 2.35 & 2.73 & 3.09 \\
\hline
\end{tabular}
\label{RSS_comparison}
\end{table}


%





\ifCLASSOPTIONcaptionsoff
  \newpage
\fi
\bibliographystyle{IEEEbib}
\bibliography{strings}

\begin{IEEEbiography}[{\includegraphics[width=1in,height=1.25in,clip,keepaspectratio]{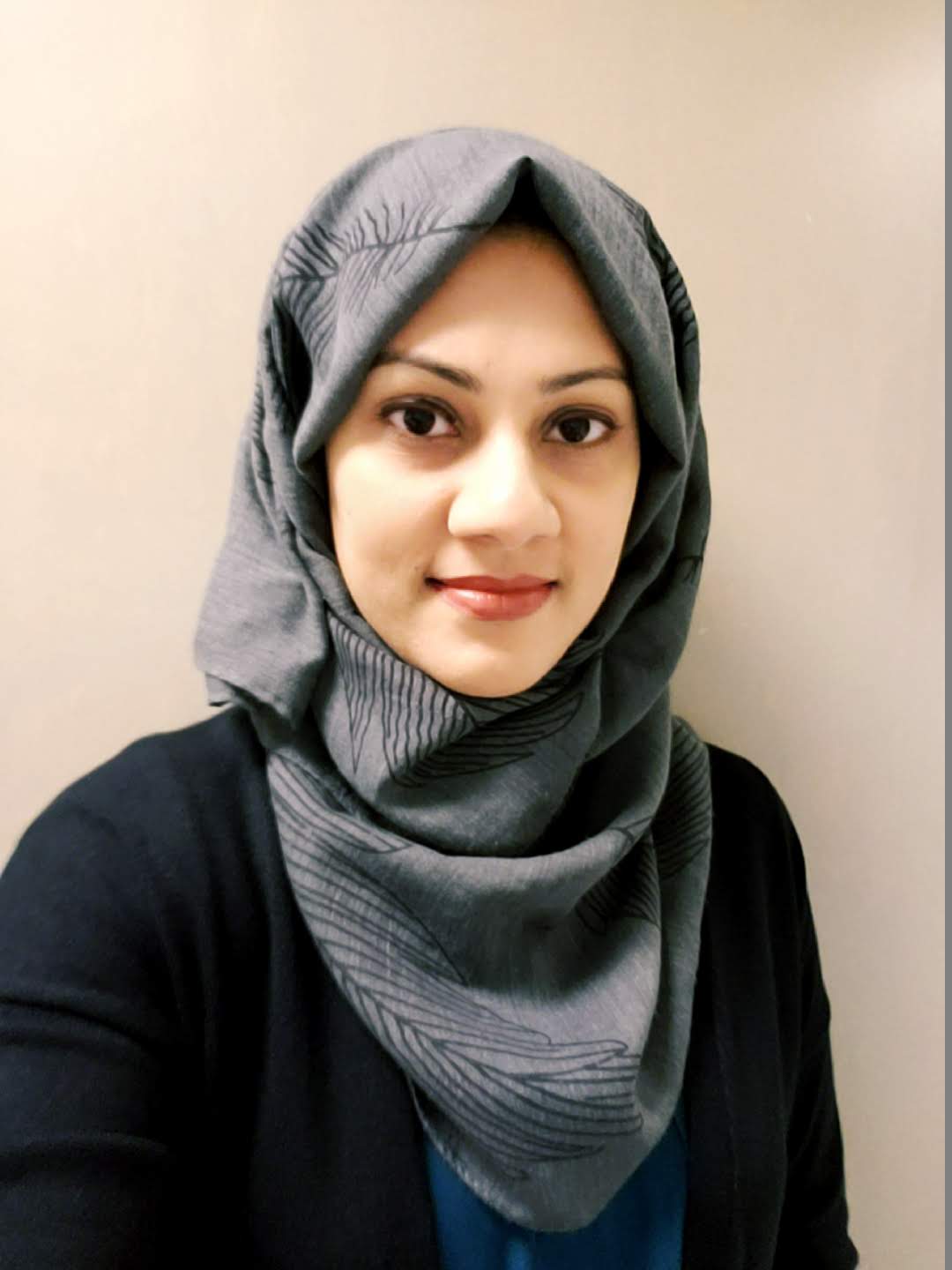}}]{Tahsina Farah Sanam} (IEEE S'10) 
received the B.Sc. and M.Sc. degrees in electrical and electronic engineering from
Bangladesh University of Engineering and Technology, Dhaka, Bangladesh, in 2009 and 2012, respectively. She is presently pursuing the PhD degree at the department of Electrical and Computer Engineering of Rutgers University, New Jersey, USA. Her research interests
include the areas of machine learning, localization and tracking; and speech signal processing. 

She won the ECE Ph.D. Research Excellence Award in Fall 2018 and School of Graduate Studies 2018-19 Conference Travel Award at Rutgers. Ms. Sanam served as the Vice Chair of Women in Engineering Affinity Group, IEEE Bangladesh Section from 2010 to 2011. Currently She is serving as the President, Society of Women Engineers Grad Chapter at Rutgers University.
\end{IEEEbiography}

\begin{IEEEbiography}[{\includegraphics[width=1in,height=1.25in,clip,keepaspectratio]{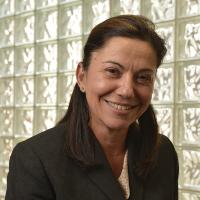}}]{Hana Godrich}
(IEEE S'91-M'04-SM'12) is an Associate Research Professor with the Electrical and Computer Engineering department at Rutgers University, Piscataway, NJ. From 2010 to 2012 she was a research scholar with Princeton University, Princeton, NJ. From 1993 until 1995, Dr. Godrich was with Scitex, Inc. (currently H-P). From 1996 until 2003, she was a partner and consultant with Enetpower, Inc., Israel, where her focus was on power systems design for mission-critical facilities. Her research interests include statistical signal processing with application to wireless sensor networks, communication, smart grid, and radar systems. 

She received the B.Sc. degree in electrical engineering from the Technion Israel Institute of Technology, Haifa, in 1987, the M.Sc. degree in electrical engineering from Ben-Gurion University, Beer-Sheva, Israel, in 1993, and the Ph.D. degrees in electrical engineering from the New Jersey Institute of Technology (NJIT), Newark, in 2010.
\end{IEEEbiography}


\end{document}